\documentclass[aps,prc,twocolumn,nofootinbib,amsmath,superscriptaddress,floatfix,10pt]{revtex4-1}
\usepackage{dcolumn}
\usepackage{graphicx}
\usepackage{mathptmx}
\usepackage{epstopdf}  
\usepackage[pdftex,colorlinks,linkcolor = blue,urlcolor  = blue,citecolor = blue]{hyperref}
\usepackage{hyperref}
\usepackage{latexsym}
\usepackage{blindtext}
\usepackage[usenames, dvipsnames]{color}
\usepackage{tabu} % to set tables to page width
\usepackage{xcolor}

\newcommand{\Os}{${}^{192}$Os} % making it faster to write Os192
\newcommand{\Osreac}{${}^{192}$Os$(\alpha$, $\alpha ' \gamma){}^{192}$Os}
\newcommand{\Osreacto}{${}^{191}$Os$(n$, $\gamma){}^{192}$Os}
\newcommand{\betadecay}{$\beta$-decay} % making it faster to write 
\newcommand{\bref}{B$^2$FH} % making it faster to write 
\newcommand{\gsf}{$\gamma$-strength function}
\newcommand{\ga}{$\gamma$}

\begin{document}

\title{First experimental constraint on the $^{191}$Os$(n,\gamma)$ reaction rate relevant to $s$-process nucleosynthesis }

\author{I.~K.~B.~Kullmann}
\email{i.k.b.kullmann@fys.uio.no}
\email{ina.kullmann@ulb.ac.be}
\affiliation{Department of Physics, University of Oslo, N-0316 Oslo, Norway}
\affiliation{Institut d'Astronomie et d'Astrophysique, Universit\'e Libre de Bruxelles, CP-226, 1050 Brussels, Belgium}
\author{A.~C.~Larsen}
\email{a.c.larsen@fys.uio.no}
\affiliation{Department of Physics, University of Oslo, N-0316 Oslo, Norway}
\author{T.~Renstr\o m}
\affiliation{Department of Physics, University of Oslo, N-0316 Oslo, Norway}
\author{K.~S.~Beckmann}
\affiliation{Department of Physics, University of Oslo, N-0316 Oslo, Norway}
\author{F.~L.~Bello Garrote}
\affiliation{Department of Physics, University of Oslo, N-0316 Oslo, Norway}
\author{L.~Crespo Campo}
\affiliation{Department of Physics, University of Oslo, N-0316 Oslo, Norway}
\author{A.~G\"orgen}
\affiliation{Department of Physics, University of Oslo, N-0316 Oslo, Norway}
\author{M.~Guttormsen}
\affiliation{Department of Physics, University of Oslo, N-0316 Oslo, Norway}
\author{J.~E.~Midtb\o}
\affiliation{Department of Physics, University of Oslo, N-0316 Oslo, Norway}
\author{E.~Sahin}
\affiliation{Department of Physics, University of Oslo, N-0316 Oslo, Norway}
\author{S.~Siem}
\affiliation{Department of Physics, University of Oslo, N-0316 Oslo, Norway}
\author{G.~M.~Tveten}
\affiliation{Department of Physics, University of Oslo, N-0316 Oslo, Norway}
\author{F.~Zeiser}
\affiliation{Department of Physics, University of Oslo, N-0316 Oslo, Norway}

\date{\today}

%%%%%%%%%%%%%%%%%%%%%%%%%%%%%%%%%%%%%%%%%%%%%%%%%%%%%%%%%%%%%%%%%%%%%%%%%%%%
\begin{abstract}

The nuclear level density and $\gamma$-decay strength of $^{192}$Os have been extracted using particle-$\gamma$ coincidence data from the $^{192}$Os($\alpha,\alpha^\prime\gamma$)$^{192}$Os reaction by means of the Oslo method. 
The level density is found to be a rather smooth function of excitation energy, approximately following the constant temperature model.
The $\gamma$-decay strength is compared to photoneutron cross-section data above the neutron separation energy, and to $E1$ and $M1$ strengths for nuclei in this mass region derived from primary transitions following neutron capture. 
Our results are in good agreement with these previous data and draw a consistent picture of the $\gamma$-strength function in the range $E_\gamma \approx 1.5-6 $ MeV. 

Using the measured nuclear level density and $\gamma$-decay strength as input to the nuclear-reaction code \textsf{TALYS}, we provide the first experimentally constrained Maxwellian-averaged cross section (MACS) for the \Osreacto\ reaction relevant to $s$-process nucleosynthesis. 
The systematic uncertainties introduced by the normalization procedure of the level density and \gsf\ were investigated and propagated to the calculated Maxwellian-averaged cross section. 
The obtained result of the Maxwellian-averaged cross section at $k_BT=30$ keV, $\langle \sigma \rangle_{n,\gamma}=1134\pm 375$ mb, is in very good agreement with the theoretical estimate provided by the KADoNiS project, giving experimental support to the adopted KADoNiS value.
Good agreement is also found with MACS values obtained from other libraries, such as TENDL-2017, ENDF/B-VII.0, and JEFF. 
\end{abstract}

%%%%%%%%%%%%%%%%%%%%%%%%%%%% SECTION %%%%%%%%%%%%%%%%%%%%%%%%%%%%%%%%%%%%%%%%%%%%%%%%

\maketitle

\section{Introduction}
\label{sec:int}

Over the last century, many scientific advances have profoundly improved our understanding of the origin, current state, and future of the universe. 
Despite this progress, our knowledge is far from complete, as many salient questions remain unsolved. One of the ``\textit{Eleven Science Questions for the New Century}'' \cite{questions2003} involves explaining the origin of the elements from iron up to uranium. 

In 1957, Burbidge \textit{et al.}~\cite{Burbidge1957} laid the very foundation of the nuclear astrophysics field in the article commonly referred to as \bref. The famous paper proposed the hypothesis that all except the lightest chemical elements were synthesized in stars by nuclear reactions, \textit{i.e.}, stellar nucleosynthesis. Independently, in the same year, Cameron \cite{Cameron1957} proposed a similar framework explaining the origin of the elements by several synthesis processes. Impressively, the two articles quite accurately describe the processes involved, and their theoretical frameworks are to a large extent still used today. 

Generally speaking, the elements up to the ``iron peak,'' \textit{i.e.}, $Z\approx 26$, are formed by charged-particle reactions during stellar burning. The two most dominant heavy-element nucleosynthesis processes are the slow ($s$) and the rapid ($r$) neutron capture processes.
The $s$-process accounts for about half of the elements heavier than iron and uses stable isotopes as stepping stones to build heavier elements. The main idea is that the neutron can easily be captured by the nucleus independent of charge, building up the heavier elements by a series of neutron captures and $\beta$ decays~\cite{Burbidge1957,Cameron1957}.
The discovery of Tc lines in red-giant stars by Merrill in 1952~\cite{Merrill1952} proved that neutron-capture processes could indeed take place in stellar environments. 

Presently, the main $s$-process is known to take place in low-mass, thermally pulsing asymptotic giant branch (AGB) stars, and the weak $s$-process is operating during the convective core He and shell C burning phases of massive stars ($M>8M_\odot$)~\cite{Kappeler2011}. 
For AGB stars, the primary source of neutrons is the $^{13}$C($\alpha,n$)$^{16}$O reaction taking place in $^{13}$C ``pockets'' at $T\approx 0.9\times 10^8$ K in between thermal pulses~\cite{Kappeler2011}. 
Here, $^{13}$C has been produced through the subsequent $^{12}$C($p,\gamma$)$^{13}$N($\beta^+\nu$)$^{13}$C reactions~\cite{Kappeler2011}. 
During thermal pulses, the $^{22}$Ne($\alpha,n$)$^{25}$Mg reaction is activated in the convective inter shell region if the temperatures exceed $2.5\times 10^8$ K~\cite{Karakas2014,Kappeler2011}. 
The $^{22}$Ne($\alpha,n$)$^{25}$Mg reaction is only active for a few years with a neutron density up to $10^{10}$ cm${}^{-3}$, while the $^{13}$C($\alpha,n$)$^{16}$O reaction operates over a timescale of about 10,000 years with neutron density of $\approx 10^6-10^8$ cm${}^{-3}$~\cite{Kappeler2011}. 
This results in two  $s$-process components in the AGB stars that produce very different abundance distributions due to the two different neutron sources~\cite{Karakas2014}.
For massive stars, the $^{22}$Ne($\alpha,n$)$^{25}$Mg neutron source is dominant, and is activated at temperatures around $3\times 10^8$ K and $1\times 10^9$ K for the convective core He and shell C burning phases, respectively~\cite{Kappeler2011}.
It is interesting to note that, in contrast to the rather robust $r$-process pattern found for neutron-star merger simulations (\textit{e.g.}, Ref.~\cite{Goriely2013,Just2015}), the $s$-process abundance yield depends strongly on parameters such as the initial mass,  the metallicity, and the mass-loss rate of the star.

To test the ability of nucleosynthesis models to provide reliable $s$-process yields, increasingly more sophisticated nuclear reaction-network simulations are invoked (see, \textit{e.g.}, Ref.~\cite{Cescutti2018}). 
As the $s$-process follows the valley of stability and has its end point at $^{209}$Bi, the reaction network mainly deals with ($n,\gamma$) and $\beta^-$-decay rates. 
To obtain the reaction rate in a stellar environment, the cross section of interest is averaged over the Maxwell-Boltzmann velocity distribution giving the so-called Maxwellian-averaged cross section (MACS), which in turn is used to derive the reaction rate. 
Traditionally,  MACS values relevant for the $s$-process are given at the energy $k_BT=30$ keV~\cite{Ratynski1988}.

For $s$-process nucleosynthesis, the typical timescale for neutron capture is much longer than the timescale for \betadecay\ of the involved unstable isotopes. 
However, at the so-called $s$-process \textit{branch points}~\cite{Ward1976}, the unstable compound nucleus created in the neutron capture reaction might live long enough to capture a neutron instead of undergoing $\beta$ decay. 
In the stellar environment, the branching depends on the \betadecay\ half-life, the neutron-capture rate, and the neutron density. 
Therefore, $s$-process branch points are of great interest as a ``diagnostic tool,'' as they can be used to derive the physical conditions at the $s$-process site~\cite{Kappeler2011}.
For the main $s$-process, known branch points include $^{95}$Zr, $^{147}$Nd, $^{169}$Nd, $^{185}$W, $^{186}$Re, and $^{191}$Os~\cite{Sonnabend2003}.

\begin{figure}[tb]
\centering
\includegraphics[clip,width=1.\columnwidth]{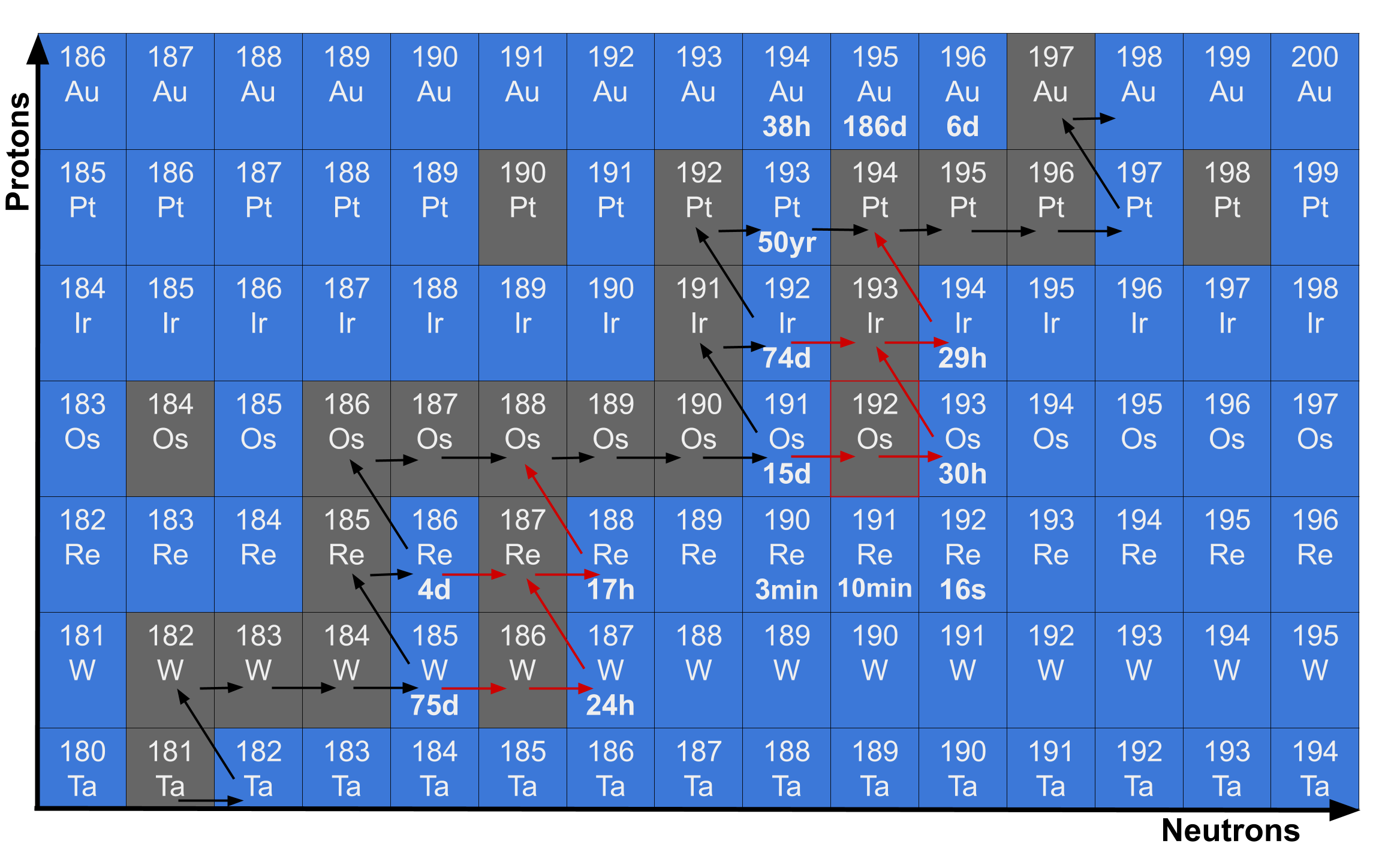}
\caption{(Color online) A section of the chart of nuclides indicating the $s$-process path with arrows. The stable (unstable) isotopes are displayed in black (blue), and a few rounded off half-lives adapted from \cite{NUDAT2} are presented in the vicinity of \Os\ and $^{186}$W.
The red arrows suggest the $s$-process path given the activation of the ${}^{185}$W and ${}^{191}$Os branch points. }
\label{fig:chart_nuclei}
\end{figure}

In \autoref{fig:chart_nuclei}, a section of the nuclear chart around the isotopes Ta, W, Re, Os, Ir, Pt, and Au is displayed. 
It is seen that the $s$-process path, indicated by the arrows, follows closely the valley of stability. 
The branch points $^{185}$W, $^{186}$Re, $^{191}$Os, $^{192}$Ir, and $^{193}$Pt are indicated with their terrestrial half-lives (note that the $\beta^+$ decay branch of $^{193}$Pt is not shown). 
The $^{191}$Os half-life is known to have a mild temperature dependence, decreasing from the terrestrial half-life of 15.4(1) days to $\simeq 8$ days at 300 MK~\cite{Diehl_2018}. 
If this branch point is activated, the neutron-capture branch may decrease the $s$-process abundances of $^{191}$Ir and $^{192}$Pt and lead to the production of $^{192}$Os, thus affecting the composition of Os and $^{193}$Ir.

%---------------------------------------------------%
%---------------------------------------------------%
%---------------------------------------------------%

For most of the ($n,\gamma$) rates of relevance for the $s$-process, neutron capture cross sections can be measured directly~\cite{Kappeler2011}. 
However, the branch point nuclei are very difficult to measure experimentally. 
There are a few cases where direct measurements of branch point nuclei have been performed~\cite{Kappeler2011}, such as for the $^{63}$Ni($n,\gamma$) reaction~\cite{Lederer2013,Weigand2015} due to the rather long half-life of $^{63}$Ni of $101.2 (15)$ years~\cite{NUDAT2}. 
For branch-point nuclei with half-lives shorter than a few days~\cite{Damone2017}, a direct experiment is extremely difficult to perform as it is challenging to produce enough sample atoms. In addition, these samples are very radioactive, so they produce a lot of background in the detectors. 
Hence, for measuring the MACS of branch-point nuclei, indirect experimental methods such as the Oslo method are in many cases the best option.

Measured $(n,\gamma)$ cross sections for the $s$-process have been presented in the compilation by Bao \textit{et al.}~\cite{Bao2000}, and a thorough evaluation is available via the KADoNiS project \cite{Dillmann2006}.
The library includes both experimental and theoretically calculated values, and has been updated several times. 
For the \Osreacto\ reaction investigated in this work, only a theoretical\footnote{The KADoNiS prediction is in fact a semi empirical value 
derived from interpolation of MACS values at $k_BT =30$ keV  of neighboring isotones; see p. 74 in Ref.~\cite{Bao2000}.}
 reaction rate is available \cite{KadonisWeb}, which is not surprising since $^{191}$Os is unstable, with a half-life of $15.4(1)$ days~\cite{NUDAT2}.  
Although this is a much shorter half-life than the important branching at $^{185}$W~\cite{Kappeler2011} with $T_{1/2}=75.1(3)$ days~\cite{NUDAT2}, it is significantly longer than, for example, the possible branchings at $^{186}$Re ($T_{1/2}= 3.7186(5)$ days~\cite{NUDAT2}) and $^{148}$Pm ($T_{1/2}= 5.368(7)$ days~\cite{NUDAT2}).
Therefore, it is possible that $^{191}$Os could be an activated $s$-process branch point, and, if so, it could impact the $s$-process yield for the heavier $s$-process elements. 
Several other theoretical and evaluated nuclear libraries are available through the JANIS project \cite{JANISWeb}, including TENDL-2017 (TALYS Evaluated Nuclear Data Library) \cite{TENDL-2017Web}, ENDF \cite{CHADWICK2006}, and JEFF \cite{JEFFWeb}.

Three nuclear input parameters are of key importance when calculating $(n,\gamma)$ cross sections: the nuclear level density (NLD), the \gsf\ (\ga SF), and the neutron optical-model potential (n-OMP) \cite{Arnould2007}. The main goal of this work is to give a first experimental constraint on the MACS value for the $^{191}$Os$(n,\gamma)$ reaction by means of an experimentally obtained NLD and \ga SF of \Os. 
The present article is part of an extensive project with the goal of experimentally constraining the MACS values of the $s$-process branch points $^{185}$W, $^{186}$Re, $^{191}$Os, and $^{193}$Ir with the Oslo method. 
We intend to perform a full $s$-process network calculation when the MACS results of the remaining isotopes are finalized.

This article is organized as follows. 
The experimental set up and data calibration will be described in \autoref{sec:exp_analysis}, while \autoref{sec:deter_nld_gsf} introduces the Oslo method utilized to extract the level density and \gsf\ of \Os. A careful uncertainty estimate of the normalization parameters of the NLD and \ga SF will be performed in \autoref{sec: sys_error_est}. In \autoref{sec:discussion}, the experimental results will be discussed, while
\autoref{sec:calc_ng} presents the Maxwellian-averaged cross section of the \Osreacto\ reaction calculated by means of the level density and \gsf\ of \Os. A summary of the main findings and an outlook are given in \autoref{sec:sum}.

%%%%%%%%%%%%%%%%%%%%%%%%%%%% SECTION %%%%%%%%%%%%%%%%%%%%%%%%%%%%%%%%%%%%%%%%%%%%%%%%

\section{Experimental setup and data analysis}
\label{sec:exp_analysis}

The experiment was performed in May 2017 at the Oslo Cyclotron Laboratory (OCL) with the goal of studying the \Osreac\ reaction. An $\alpha$ beam with an energy of 30 MeV was applied with a beam current of  $\approx 4$ nA for three days. The \Os\ target was self supported, $> 99$\% enriched, and with a thickness of 0.33 mg/cm$^2$. 
A self-supporting ${}^{60}$Ni target with a thickness of 2 mg/cm$^2$ was placed in the beam for approximately 2 hours at the beginning of the experiment for calibration purposes. 
 
The CACTUS/SiRi detector array is designed to study particle-$\gamma$ coincidences.
The SiRi detector array measures charged particles emitted in the solid angle that the array covers ($\approx 6$\% of 4$\pi$), and consists of eight separate silicon detectors in a ring placed inside the target chamber. Each silicon detector is divided into two parts, a thin $\Delta E$ (130 $\mu$m) and a thick $E$  (1550 $\mu$m) detector \cite{Guttormsen2011}. Each of the eight $\Delta E$ detectors are divided into eight strips, forming a $\Delta E$-$E$ particle-telescope system of 64 detectors. 
In the current experiment, the SiRi array was placed in backward angles, covering angles from 126$^\circ$ to 140$^\circ$ between the incident trajectory and the trajectory of the emitted particle \cite{Guttormsen2011}. The angular resolution of the strips are $\approx 2^\circ$ \cite{Guttormsen2011,CrespoCampo2016} and the experimental resolution of the telescopes was observed to be $\approx 235$ keV full width at half maximum for the \Osreac\ reaction. 

The energies of the coincident $\gamma$ rays were measured by the CACTUS detector system~\cite{CACTUS} consisting of 26 collimated NaI(Tl) scintillation detectors.
Each cylindrically-shaped crystal is $5$ \textit{in.} $\times 5$ \textit{in.} in size and was mounted on a spherical frame surrounding the target chamber giving a total measured efficiency of 14.1(1)\% at $E_\gamma=1332.5$ keV, as measured with a $^{60}$Co source at a distance of 22 cm from the center of the target to the front of the NaI crystals~\cite{CrespoCampo2016}. 
The full width at half maximum of CACTUS is measured to the 6.8\% at $E_\gamma=1332.5$ keV.
More information on the detector and data-acquisition electronics setup at OCL is given in Refs.~\cite{Larsen2008,Guttormsen2011}.

For the energy calibration of the SiRi particle detector array, we need two reference points sufficiently separated in energy in the $\Delta E$-$E$ plot. 
Due to the reaction kinematics, we chose the elastic peak of ${}^{60}$Ni in addition to the elastic peak of \Os. 
As an example, for the 126$^\circ$ angle, the deposited energies in the $\Delta E$ detectors are 7.56 and 6.34 MeV for the $^{60}$Ni and \Os\ ground states, respectively, while the corresponding, deposited energies in the $E$ detectors are 15.2 and 20.9 MeV. 
As reference peaks for the CACTUS $\gamma$ detector we use (1) the transition from the first excited $2^+$ state to the ground state of $^{60}$Ni at 1332.5 keV in the calibration data set, and (2) the transition from the $1/2^+$ excited state to the ground state of the contaminant $^{15}$N at 5298.8 keV from the ${}^{12}$C$(\alpha,p){}^{15}$N reaction.

\begin{figure*}[tb]
\centering
\includegraphics[clip,width=1.\textwidth]{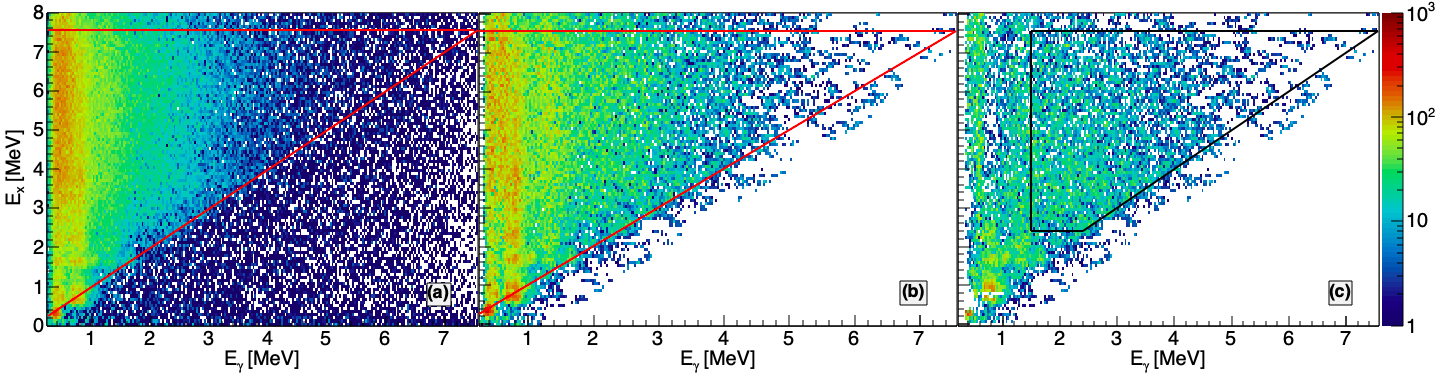}
\caption{(Color online) Excitation energy $E_x$ versus $\gamma$ energy $E_\gamma$ for \Os: \textbf{(a)} the original, \textbf{(b)} unfolded, and \textbf{(c)} first-generation matrices. The number of counts are represented by the color scale and the bin widths (see the text) are 32 keV and 80 keV for the $x$ and $y$ axes, respectively. In \textbf{(a)} and \textbf{(b)} the diagonal line $E_x = E_\gamma$ is displayed in red, and in \textbf{(c)} the black solid lines represent the limits included in the $\chi^2$-minimization method.
Note that the bin values below the diagonal in \textbf{(a)} fluctuate around zero with counts $ <10$. Before unfolding, all data points in the region below the diagonal plus the resolution of the NaI(Tl) detector $\delta E_\gamma$, \textit{i.e.}, counts below the line $E_x = E_\gamma + \delta E_\gamma$, were set to zero.
 }
\label{fig:coincidence}
\end{figure*}

The SiRi particle telescope allows for particle identification by measuring the deposited energy of the ejected charged particles in the thin $\Delta E$ and thick $E$ detectors. The particles' difference in charge and mass separate them in the $\Delta E$-$E$ matrix. Therefore, a two-dimensional gate on the emitted $\alpha$ particles in the $\Delta E$-$E$ matrix was set to select the data corresponding to the $(\alpha,\alpha'\gamma)$ reaction. 
By using the known reaction kinematics, the deposited energy in the SiRi detector, \textit{i.e.}, the $\Delta E$-$E$ spectra, was converted into excitation spectra $E_x$ for the final nucleus. 
Then, the coincidence events were sorted into a matrix with the excitation energy $E_x$ versus the $\gamma$ energy $E_\gamma$ on the $y$ and $x$ axes, and the number of counts on the $z$ axis. Every point in the coincidence matrix corresponds to the simultaneous detection of an $\alpha$-particle and $\gamma$ ray(s) related to the \Osreac\ reaction; see \autoref{fig:coincidence}\textcolor{blue}{a}.

In general, our choice of bin widths for the $E_x$ and $E_\gamma$ axes is connected to the resolution of the SiRi and CACTUS detectors, and also to the statistics. For the low-energy $\gamma$ rays, we have a gradual threshold starting at about 350-400 keV due to the chosen range for the Analog-to-Digital Converters (ADCs) (up to about 14 MeV), and the CACTUS resolution goes from about 50 keV full width at half maximum for 600 keV $\gamma$ rays to about 250-270 keV full width at half maximum for 8 MeV $\gamma$ rays. To ensure a sufficient number of channels in $\gamma$ ray peaks with energies down to the threshold (peaks with resolution of 40-50 keV full width at half maximum), we use a bin width of 8 keV for the $\gamma$ rays initially in the analysis. This small bin width for the $\gamma$ rays is kept throughout the two first steps of the analysis, \textit{i.e.}, the unfolding and the first-generation method, and is increased to $>100$ keV for the extraction of the level density and gamma strength to ensure sufficient statistics in each $E_x$-$E_\gamma$ “pixel”. Similarly, for the excitation energy, we use initially a smaller bin width of 8 keV. However, for the plots of \autoref{fig:coincidence}, we chose to re bin to 32 and 80 keV on the $E_x$ and $E_\gamma$ axes, respectively, to make the plots more informative (more counts in each “pixel”).

For each excitation-energy bin, the \ga-ray spectra were unfolded \cite{Guttormsen1996} with respect to the response of the detectors. The method developed in \cite{Guttormsen1996} has proved to work well for continuum \ga\ rays and preserves the experimental statistical uncertainties without introducing any artificial fluctuations. In Table 1 of \cite{CrespoCampo2016}, the applied updated response functions from 2012 are shown. The resulting unfolded coincidence matrix for \Os\ is presented in \autoref{fig:coincidence}\textcolor{blue}{b}. 

%---------------------------------------------------%
%---------------------------------------------------%
%---------------------------------------------------%
 
To obtain the transition probabilities a first-generation, or primary, $\gamma$-ray spectrum must be identified. Starting from the unfolded spectrum, the first-generation spectra are obtained by an iterative subtraction method \cite{Guttormsen1987}. 
This method relies on the assumption that the decay routes of a $\gamma$ ray from a level at an excitation-energy bin is independent of how the levels within the bin were reached. Then, the shape of the $\gamma$-ray spectra involving states populated by the nuclear reaction is identical to the spectra created by states populated by higher-lying energy states. \autoref{fig:coincidence}\textcolor{blue}{c} shows the first-generation matrix, or the decay probability of \Os\ for a given excitation energy $E_x$ and $\gamma$-ray energy $E_\gamma$. A detailed discussion of the uncertainties and errors of the unfolding and first-generation method can be found in \cite{Larsen2011}.

The next step in the analysis is to obtain the functional form of the NLD and the \ga SF by means of the iteration procedure described in \cite{Schiller2000}. 
The \ga-ray transmission coefficient $\mathcal{T}$ is related to the strength function $f$ through \cite{Kopecky1990}:
\begin{equation}
f_{XL} (E_\gamma) = \frac{1}{2\pi} \frac{\mathcal{T}_{XL} (E_\gamma)}{E^{(2L+1)}_\gamma }.
\label{eq:f_to_trans}
\end{equation}
where $E_{\gamma}$ is the $\gamma$-ray energy with electromagnetic character $X$ and multipolarity $L$. 
This relation is obtained by combining the expression for the $\gamma$-ray transmission coefficient following Blatt and Weisskopf~\cite{Blatt_Weisskopf},
\begin{equation}
    \left<\Gamma^{XL}_\gamma(E_x,E_\gamma,I,\pi)\right> = \mathcal{T}^{XL}_{\gamma}(E_x,E_\gamma,I,\pi) \frac{D(E_x,E_\gamma,I,\pi)}{2\pi},
\label{eq:transdef}
\end{equation}
where $\left< \Gamma^{XL}_\gamma(E_x,E_\gamma,I,\pi)\right>$ is the partial radiative width from excitation energy $E_x$ and from levels with spin, and parity $I$ and $\pi$, and $D$ is the average level spacing, with the definition of the \ga SF given by Bartholomew \textit{et al.}~\cite{Bartholomew1973}:
\begin{equation}
    {f}_{XL}(E_x, I, \pi, E_\gamma)= \frac{\left<\Gamma^{XL}_\gamma(E_x, I, \pi, E_\gamma)\right>}{D(E_x,E_\gamma,I,\pi) E_\gamma^{2L+1}}. 
\label{eq:gsfdef}
\end{equation}
By application of the Brink hypothesis~\cite{Brink1955}, the dependence on $E_x,I,\pi$ is averaged out and the $\gamma$-ray transmission coefficient (and correspondingly the \ga SF) is dependent only on the $\gamma$-ray energy, giving \autoref{eq:f_to_trans}.
As is common practice, we assume dipole radiation ($L=1$) to be the main contribution to the experimental $f$, as supported by experimental data (\textit{e.g.}, Refs.~\cite{Kopecky2017,Larsen2013}).

If a statistical decay process\footnote{The reaction leads to a compound nucleus which decays independently of how it was formed \cite{Bohr1936}. }
is assumed, and by applying Fermi's golden rule \cite{Guttormsen2011b}, the experimental first-generation matrix $P(E_x, E_\gamma) $ can be factorized into the nuclear level density $\rho$ and the \ga-ray transmission coefficient $\mathcal{T}$. In addition, the factorization assumes the Brink hypothesis~\cite{Brink1955}
to be valid,\footnote{See Refs.~\cite{Larsen2011,Guttormsen2016} for a thorough discussion of the validity of the Brink hypothesis when applied in the Oslo method. } \textit{i.e.} it is assumed that the \ga-ray transmission coefficient is independent of excitation energy. We therefore write the factorization of the experimental first-generation matrix as~\cite{Schiller2000}
\begin{equation}
P(E_x, E_\gamma) = \frac{ \mathcal{T}(E_\gamma) \rho(E_x - E_\gamma) }{ \sum_{E_\gamma = E_\gamma^{\mathrm{min}}}^{E_x} \mathcal{T}(E_\gamma) \rho(E_x - E_\gamma)   },
\label{eq: first_gen_propto}
\end{equation}
where $\rho(E_x - E_\gamma)$ is the level density at the final energy level after the emission of a \ga\ ray of energy $E_\gamma$ at excitation energy $E_x$, and $\mathcal{T}$ is the \ga-ray transmission coefficient. 
The first-generation matrix is normalized so that the sum over all $\gamma$-ray energies in the range $E_\gamma^{\mathrm{min} } \leq E_\gamma \leq E_\gamma^{\mathrm{max}}$ is unity, for a given excitation energy bin $E_x$:
\begin{equation}
\sum_{E_\gamma = E_\gamma^{\mathrm{min}}}^{E_x} P(E_x, E_\gamma) = 1.
\end{equation}
Here, $E_\gamma^{\mathrm{min}}$ and $E_\gamma^{\mathrm{max}}$ are the minimum and maximum $\gamma$-ray energies included in the analysis. The limits are chosen to ensure that the data employed stem from the statistical excitation energy region in the first-generation matrix $P(E_x,E_\gamma)$. In the present work, the upper and lower limits of the matrix were set to $E_\gamma^{\mathrm{min}}=1.5$ MeV and $E_x^{\mathrm{min}}=2.4$ MeV, while the maximum excitation energy was set to the neutron separation energy $E_x^{\mathrm{max}}=7.558(2)$ MeV (recently re measured by Wang \textit{et al.}~\cite{Wang2017}) to exclude the neutron emission channel. 
The section of the first-generation matrix $P$ that was included in the analysis is contained by the black solid lines in \autoref{fig:coincidence}\textcolor{blue}{c}. Given a reasonable choice for the $E_\gamma^{\mathrm{min}}$ limit, higher and lower values of $E_x^{\mathrm{min}}$ do not change the results significantly, \textit{i.e.}, the extraction of $\rho$ and $\mathcal{T}$ is not sensitive to the choice of $E_x^{\mathrm{min}}$ within the experimental error bars for \Os, when $E_\gamma^{\mathrm{min}}$ is held constant.

The \textit{functional form} of $\rho$ and $\mathcal{T}$ for \Os\ was then uniquely determined by a $\chi^2$ minimization \cite{Schiller2000} of \autoref{eq: first_gen_propto}. The minimization method is only required to fulfill \autoref{eq: first_gen_propto}, no other assumptions of the functional form of $\rho$ and $\mathcal{T}$ were done. 
For the extraction of the level density and the transmission coefficient, the experimental first-generation matrix is re-binned to 160 keV per bin for the $E_x$ and $E_\gamma$ axis to ensure enough statistics for the $\chi^2$-minimization method. 
The remaining task is then to normalize the NLD and the \ga SF to experimental data.

%%%%%%%%%%%%%%%%%%%%%%%%%%%% SECTION %%%%%%%%%%%%%%%%%%%%%%%%%%%%%%%%%%%%%%%%%%%%%%%%

\section{Determining the nuclear level density and the \gsf} 
\label{sec:deter_nld_gsf}

If one solution $\mathcal{T}(E_\gamma)$ and $\rho(E_f)$ of \autoref{eq: first_gen_propto} is known, it can be mathematically shown \cite{Schiller2000} that there exists an infinite set of solutions given by the transformation
\begin{align}
\widetilde{\rho}(E_x - E_\gamma) &= A \exp[\alpha(E_x - E_\gamma)]\ \rho(E_x - E_\gamma), \\
\widetilde{\mathcal{T}}(E_\gamma) &= B \exp[\alpha E_\gamma]\ \mathcal{T}(E_\gamma). \label{eq: trans_coeff_sol}
\end{align}
Any combination of values $A$, $B$ and $\alpha$ will yield solutions $\widetilde{\rho}$ and $\widetilde{\mathcal{T}}$ obeying
\begin{equation}
P(E_x, E_\gamma) = \frac{ \widetilde{\mathcal{T}}(E_\gamma) \widetilde{\rho}(E_x - E_\gamma) }{\sum_{E_\gamma = E_\gamma^{min}}^{E_x} \widetilde{\mathcal{T}}(E_\gamma) \widetilde{\rho}(E_x - E_\gamma)   }.
\label{eq: rhosig_general}
\end{equation}
The solutions $\rho$ and $\mathcal{T}$ obtained in the $\chi^2$-minimization procedure are therefore the general solution of \autoref{eq: first_gen_propto}: they contain the general shape and the functional form of the solution. The physical, or special solution is determined by the parameters $A$, $\alpha$, and $B$. These parameters are acquired separately for the NLD and \ga SF by normalizing the functions to external data. 

None of the required experimental parameters for the normalization of the NLD and \ga SF are available for \Os. Therefore, these parameters have to be estimated through systematics, which introduce an additional uncertainty in the NLD and \ga SF for \Os. 

%%%%%%%% SUBSECTION %%%%%%%%
\subsection{Normalization of the level density}
\label{subsec:norm_nld}

The absolute value $A$ and the slope $\alpha$ of the level density $\rho$ are determined by (1) a normalization to the number of known discrete levels at low excitation energy, and (2) the estimated level density at the neutron separation energy, $\rho(S_n)$. 

To calculate $\rho(S_n)$ from neutron resonance data ($D_0$ values), the spin and parity distributions of the level density at $S_n$ have to be known. These quantities are model dependent in this case (for some nuclei this is experimentally known), introducing a potentially large uncertainty in the normalization procedure. 

In this work, only the constant-temperature (CT) formula is considered. The CT model provides a simple, analytic formula for the NLD:
\begin{equation}
\rho_{CT}(E) = \frac{1}{T_{CT}} \exp \Bigg( \frac{E - E_0}{T_{CT}} \Bigg),
\label{eq: rho_CT}
\end{equation}
where the nuclear temperature $T_{CT}$ and the energy shift $E_0$ serve as free parameters to be adjusted to the experimental discrete levels. The CT formula assumes an equiparity distribution, \textit{i.e.}, that both parities contribute equally to the level density. 
The spin distribution is approximated by \cite{Ericson1960}
\begin{equation}
g(E_x,I) \simeq \frac{2I + 1}{2\sigma_I^2(E_x)} \exp[-(I+1/2)^2/2\sigma_I^2(E_x)],
\end{equation}
for a specific excitation energy $E_x$ and spin $I$. 
This expression is valid within the statistical model assuming random couplings of angular momenta, provided that many particles and holes are excited~\cite{Ericson1960}. 
Following the approach of \cite{Guttormsen2017}, the energy-dependent spin cutoff parameter is introduced as
\begin{equation}
\sigma_I^2(E_x) = \sigma_d^2 + \frac{E_x - E_d}{S_n - E_d}[\sigma_I^2(S_n) - \sigma_d^2],
\label{eq: spin_cutoff}
\end{equation}
going through two anchor points $\sigma_d^2$ and $\sigma_I^2(S_n)$ at the energies $E_d$ and $S_n$, respectively. In this approach, the $n$ known discrete levels at the excitation energy $E_x=E_d=1.1$ MeV are used to estimate an experimental spin distribution to determine the first point $\sigma_d^2$. The energy of $E_d$ is chosen at the highest possible energy, before most of the experimental spin assignments become uncertain. For \Os, several spin assignments are uncertain from $1362$ keV \cite{NNDC}. Therefore, we choose the excitation-energy centroid of the experimental spin distribution to be at $E_x=1.1$ MeV, where the level scheme is regarded complete. 
Assuming a rigid moment of inertia $\Theta=0.0146A^{5/3}$, the second point at $E_x=S_n=7.558$ MeV can be estimated through  \cite{VonEgidy2005}
\begin{equation}
\sigma_I^2(S_n) = 0.0146A^{5/3} \frac{1 + \sqrt{1 + 4aU_n}}{2a},
\label{eq: sig_sn}
\end{equation}
where $A$ is the mass number and $U_n=S_n - E_{sh}$ is the intrinsic excitation energy. The level parameter $a$ and the energy shift parameter $E_{sh}$ are taken from \cite{VonEgidy2005}. The applied parameters to obtain the spin-cutoff parameter in this work can be found in \autoref{tab:rho_paraCT}. 

%---------------------------------------------------%
\begin{table}[tb]
\centering
\caption{Parameters used to extract the level density within the CT model approach, in addition to the level-density parameter $a$ and back shift $E_{sh}$ for the Fermi-gas model from \cite{VonEgidy2005}. }
\begin{tabular}{lccccr}
\hline \hline
$a$ [MeV$^{-1}$] & $E_{sh}$ [MeV] & $E_d$ [MeV] & $\sigma_d$ & $T_{CT}$ [MeV] & $E_0$ [MeV] \\ 
  &  					 	 & 			  			 & 						 & 							  & 	\\ 
\hline
 18.472 					  & 0.328			 & 1.1 					  & 2.8 				& 0.5 						   & 0.331 \\ 
\hline \hline
\end{tabular}
\label{tab:rho_paraCT}
\end{table}
%---------------------------------------------------%
%---------------------------------------------------%

We can then estimate the level density at the neutron separation energy $S_n$ from the spacing of neutron s-wave ($\ell=0$) resonances $D_0$ following $(n,\gamma)$ capture \cite{Schiller2000}:
\begin{equation}
\rho(S_n) = \frac{2\sigma_I^2}{D_0} \frac{1}{(I_t+1) \exp{[-(I_t+1)^2/2\sigma_I^2}] + I\exp{[-I_t^2/2\sigma_I^2]}}.
\label{eq: rho_sn}
\end{equation}
Here, we implement the spin cutoff parameter $\sigma_I = \sigma_I(S_n)$ for the compound nucleus following neutron capture from \autoref{eq: spin_cutoff}, and $I_t$ is the ground-state spin of the target nucleus in the $(n,\gamma)$ reaction, \textit{i.e.}, $9/2^-$ for ${}^{191}$Os. 

%---------------------------------------------------%
\begin{figure}[tb]
\centering
\includegraphics[clip,width=1.\columnwidth]{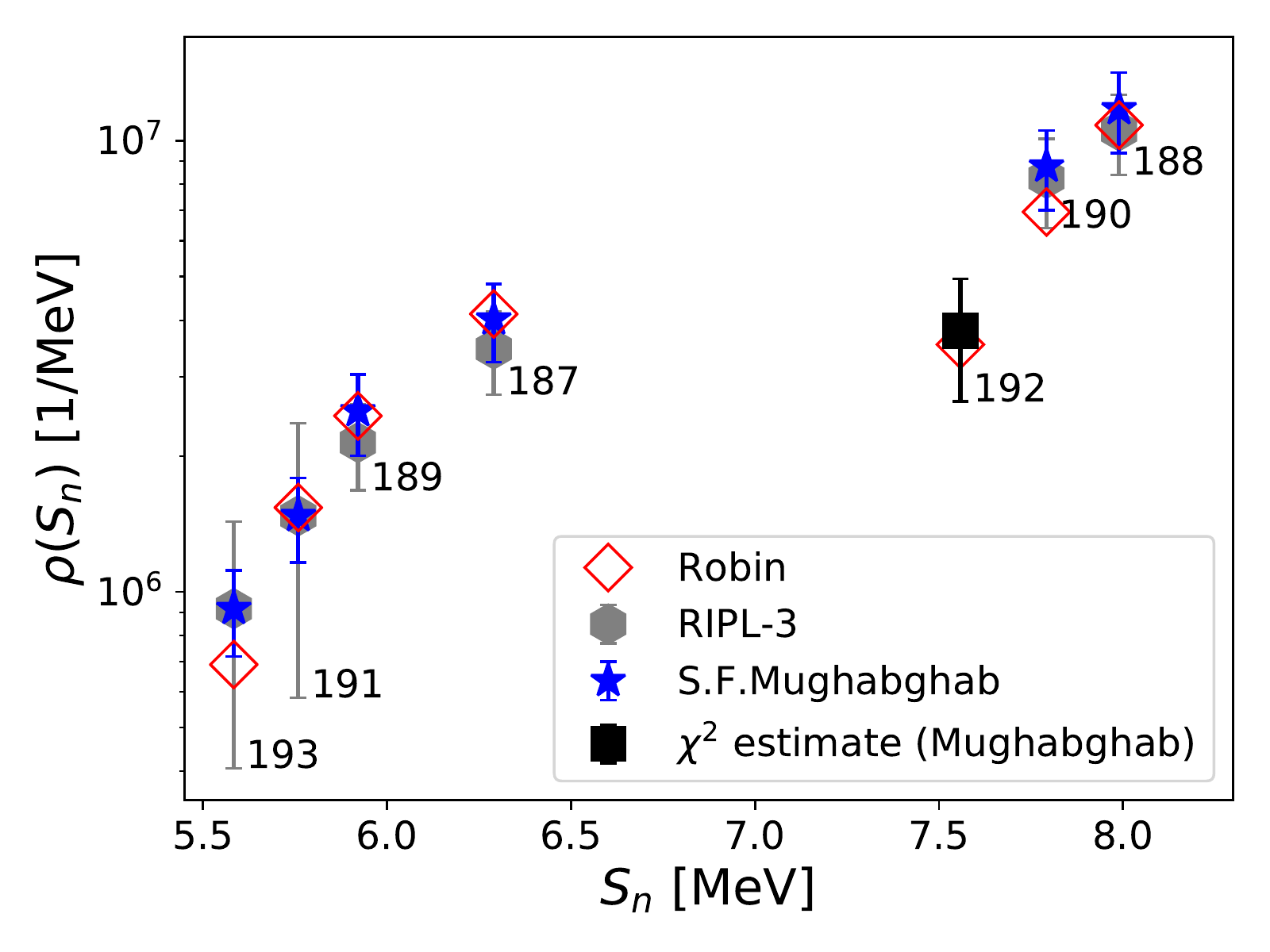}
\caption{(Color online) The level density at the separation energy $\rho(S_n)$ plotted against the neutron separation energy $S_n$. The mass numbers $A$ are annotated next to the data points. The values are calculated from evaluated experimental $D_0$ values (RIPL-3 \cite{Capote2009}, Mughabghab \cite{Mughabghab2006}) using spin cutoff parameters according to EB05/06 \cite{VonEgidy2005}.  }
\label{fig:D0_sys}
\end{figure}
%---------------------------------------------------%
%---------------------------------------------------%

In order to normalize the level density of \Os, the $D_0$ parameter of \autoref{eq: rho_sn} must be known. Unfortunately, no tabulated $D_0$ values are available for ${}^{191}$Os for the NLD of \Os. Therefore, an estimate is obtained by using available values from neutron resonance experiments of isotopes in the same mass region. In \autoref{fig:D0_sys}, $\rho(S_n)$ values for all available osmium isotopes are presented. The evaluated experimental $D_0$ values for  $A=187-193$ (except 192) are taken from \cite{Mughabghab2006}\footnote{The new, updated version of this atlas \cite{Mughabghab2018} (available February, 2018) was not included in this work. The impact of the few updated $D_0$ and $\langle \Gamma_{\gamma0} \rangle$ values on the recommended $\rho(S_n)$ and $\langle \Gamma_{\gamma0} \rangle$ for \Os\ was tested. The recommended values only changed by 5\%, an insignificant adjustment compared to the applied uncertainties of $ 20-40$\%. }
 (blue star) and \cite{Capote2009} (gray hexagon) and transformed into $\rho(S_n)$ values by means of \autoref{eq: rho_sn}. The theoretical, or systematic, values (red diamond) are provided by the phenomenological parametrization of Von Egidy and Bucurescu, hereafter named EB05/06~\cite{VonEgidy2005}.
 
%---------------------------------------------------%
\begin{figure}[tb]
\centering
\includegraphics[clip,width=1.\columnwidth]{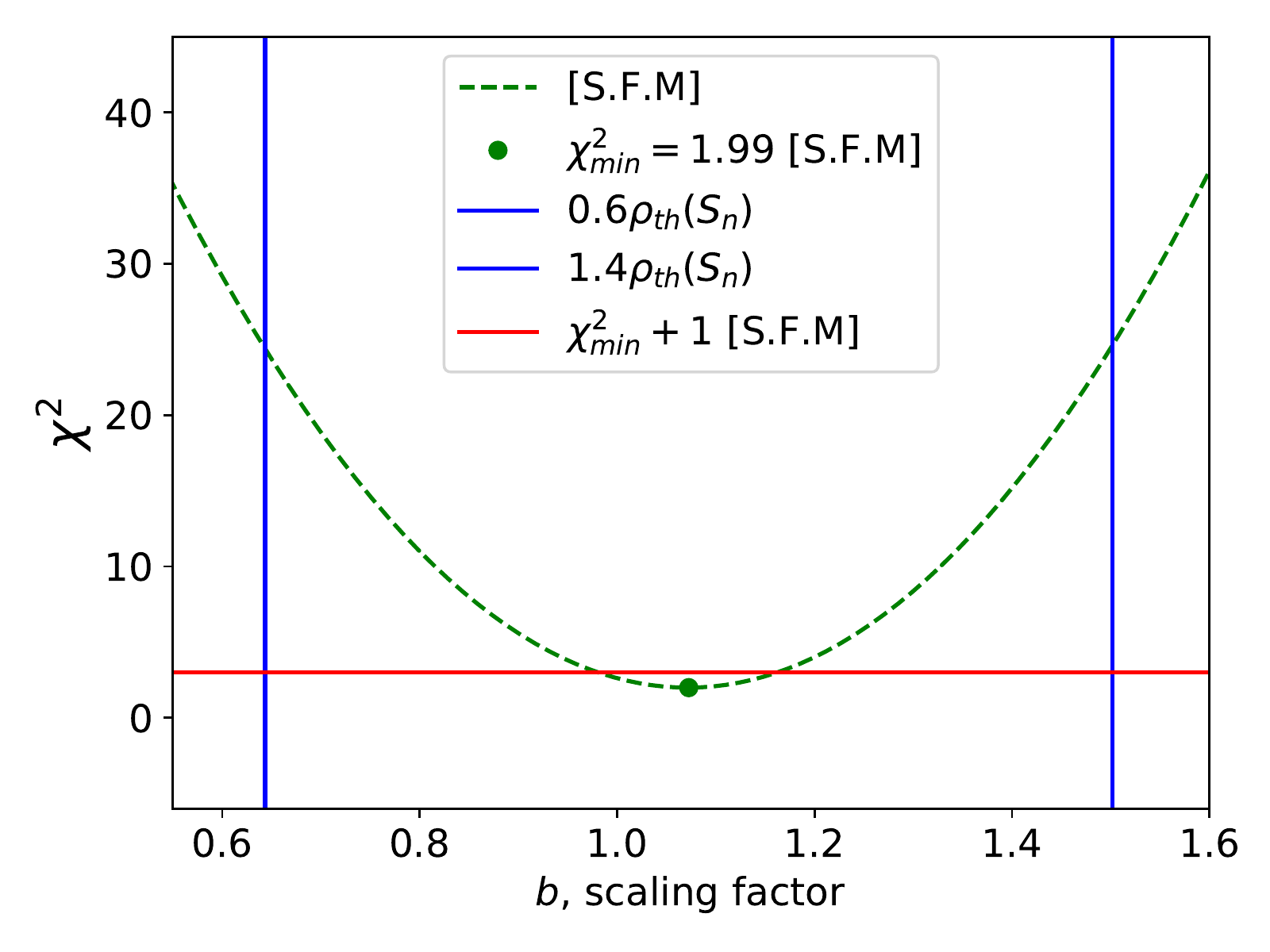}
\caption{(Color online) The scaling factor $b$ versus the $\chi^2$ value found by using evaluated experimental values from Mughabghab \textit{et al.} (S.F.M) \cite{Mughabghab2006} (green); see the text. The red line represents $\chi^2_{\mathrm{min}}+1$, and the blue lines present the $b$ values corresponding to $0.6 \rho_{th}(S_n)$ and $1.4 \rho_{th}(S_n)$.}
\label{fig:chisquare}
\end{figure}
%---------------------------------------------------%
%---------------------------------------------------%

The EB05/06 values presented in \autoref{fig:D0_sys} are generally in good agreement with the available data in this mass region. Although the general trend for the $D_0$ values are reproduced by the global fit of the EB05/06 parametrization, 
this is not a guarantee that the value for \Os\ is well reproduced. Therefore, a $\chi^2$ minimization was done to estimate the level density at the neutron separation energy for \Os\ by using the experimental values provided by \cite{Mughabghab2006}. The appropriate scaling factor $b$ between the EB05/06 \cite{VonEgidy2005} values $\rho_{\mathrm{th}}(S_n)$ and the values provided by \cite{Mughabghab2006} was found by minimizing
\begin{equation}
\chi^2 = \sum_A \frac{ \big(b \rho_{\mathrm{th}}(S_n) - \rho_{\mathrm{exp}}(S_n) \big)^2}{\Delta \rho_{\mathrm{exp}}^2(S_n)}
\end{equation}
where $\Delta \rho_{\mathrm{exp}}(S_n)$ is the listed one-standard-deviation uncertainty for the experimental $D_0$ points \cite{Mughabghab2006} and the sum runs over $A\in[187,188,189,190,191,193]$. The resulting scaling factor versus the $\chi^2$ value is presented in \autoref{fig:chisquare}, where the minimum values at $b=1.073$ and $\chi^2=1.99$ are highlighted. A rough error estimate, represented by the blue lines in \autoref{fig:chisquare}, of the resulting $\rho(S_n)$ point for \Os\ (blue in \autoref{fig:D0_sys}) was done by taking the largest uncertainty of the experimental points \cite{Mughabghab2006} and adding a 40\% relative error. Considering the spread of $D_0$ values in the literature, 40\% is a conservative estimate chosen to overestimate the uncertainty rather than underestimate it. This error analysis leads to the lower (L), recommended (R), and higher (H) estimates for the level density at the neutron separation energy listed in \autoref{tab:rec_LH}.

Following \cite{Guttormsen2017}, a systematic error band can be created by multiplying the rigid moment of inertia $\Theta=0.0146A^{5/3}$ of \autoref{eq: sig_sn} with a factor $\eta$. In this way, the lower (L), recommended (R), and higher (H) estimates of the spin cutoff parameter listed in \autoref{tab:rec_LH} can be introduced by letting $\eta \in [0.8,0.9,1.0]$, respectively.

\begin{figure}[tb]
\centering
\includegraphics[clip,width=1.\columnwidth]{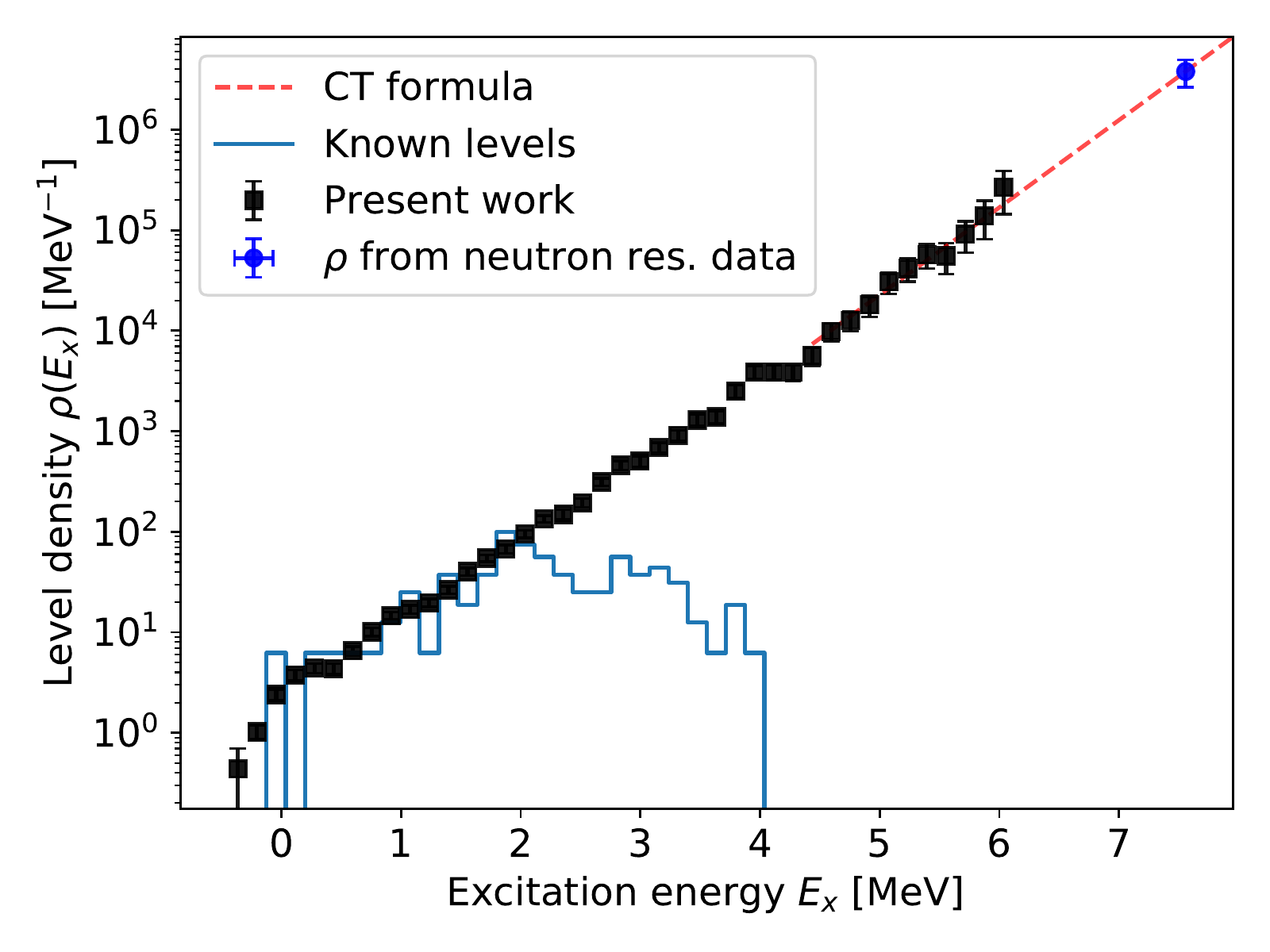}
\caption{(Color online) The level density versus excitation energy for the Oslo data (black). The CT interpolation between the data and $\rho(S_n)$ (blue) is indicated by a red line in addition to the known levels (light blue). }
\label{fig:rho_norm}
\end{figure}

In \autoref{fig:rho_norm}, the level density (black) obtained by using the recommended parameters listed in \autoref{tab:rec_LH} is presented. 
The experimental data only reach up to $\approx S_n - 1.5$ MeV, so an interpolation between the Oslo data and $\rho(S_n)$ (blue) is done using the CT level density model (red). The slope of the  experimental level density is found by forcing the level density to fit the discrete levels at $E_x=116$ keV and $E_x=1236$ keV, in addition to the CT interpolation between $E_x=3156$ keV and $E_x=6036$ keV. 
This procedure ensures a good reproduction of the cumulative number of levels up to $E_x \approx 2$ MeV. 
The CT model parameters of \autoref{eq: rho_CT} are listed in \autoref{tab:rho_paraCT}.
The uncertainties presented by the black data points are statistical errors in addition to systematic errors from the unfolding and first generation method, estimated with the Oslo method software \cite{oslo-method-software} as described in \cite{Schiller2000}, from now on referred to as Oslo method errors.

%---------------------------------------------------%
\begin{table}[tb]
\centering
\caption{The high (H), recommended (R), and lower (L) estimates of the parameters used in the normalization procedure of the level density and the $\gamma$-strength function. Note that the low (high) $D_0$ values correspond to the the high (low) $\rho(S_n)$ values.}
\begin{tabular}{ccccc} 
\hline \hline
  & $D_0$ [eV] & $\rho(S_n)$ [$10^6$ MeV$^{-1}$] & $\sigma_I(S_n)$ & $\langle \Gamma_{\gamma0} \rangle$ [meV] \\ 
\hline 
H & 5.25 & 4.94 & 7.81 & 107 \\ 
R & 3.66 & 3.79 & 7.41 & 79 \\ 
L & 2.81 & 2.64 & 6.98 & 61 \\ 
\hline \hline
\end{tabular}
\label{tab:rec_LH}
\end{table}
%---------------------------------------------------%
%---------------------------------------------------%

%---------------------------------------------------%
%---------------------------------------------------%
%---------------------------------------------------%

%%%%%%%% SUBSECTION %%%%%%%%
\subsection{Normalization of the \gsf}
\label{subsec:norm_gsf}

The shared slope $\alpha$ of the level density $\rho$ and the transmission coefficient $\mathcal{T}$ was found through the normalization of $\rho$ in the previous subsection. The parameter $B$, which gives the absolute normalization of $\mathcal{T}$, is the only remaining unknown of \autoref{eq: trans_coeff_sol}. This parameter can be constrained by the known average total radiative width, $\langle \Gamma_{\gamma0} \rangle$, at $S_n$ from $s$-wave neutron resonance experiments.
For $s$-wave neutron capture on an even-even target leading to levels with spin-parity $1/2^+$ in the compound nucleus, and using Eq.~(3.1) in Ref.~\cite{Kopecky1990}, one obtains
\begin{align}
\langle \Gamma_{\gamma0} (S_n,I^\pi=1/2^+) \rangle = &\frac{D_0(S_n,I^\pi=1/2^+)}{2\pi } \times\nonumber\\
& \sum_{XL}\sum_{I_f} \int_0^{S_n} \mathrm{d}E_\gamma \mathcal{T}_{XL}(E_\gamma) \rho(S_n-E_\gamma,I_f).
\label{eq:Gg_int}
\end{align}
The summation and integration runs over all final levels with spin $I_f$ that are accessible by $E1$ and $M1$ transitions with energy $E_\gamma$, \textit{i.e.}, assuming dipole radiation ($L=1$). Then, the transmission coefficient $\mathcal{T}$ can be transformed into the \gsf\ $f$ by using \autoref{eq:f_to_trans}.

\begin{figure}[tb]
\centering
\includegraphics[clip,width=1.0\columnwidth]{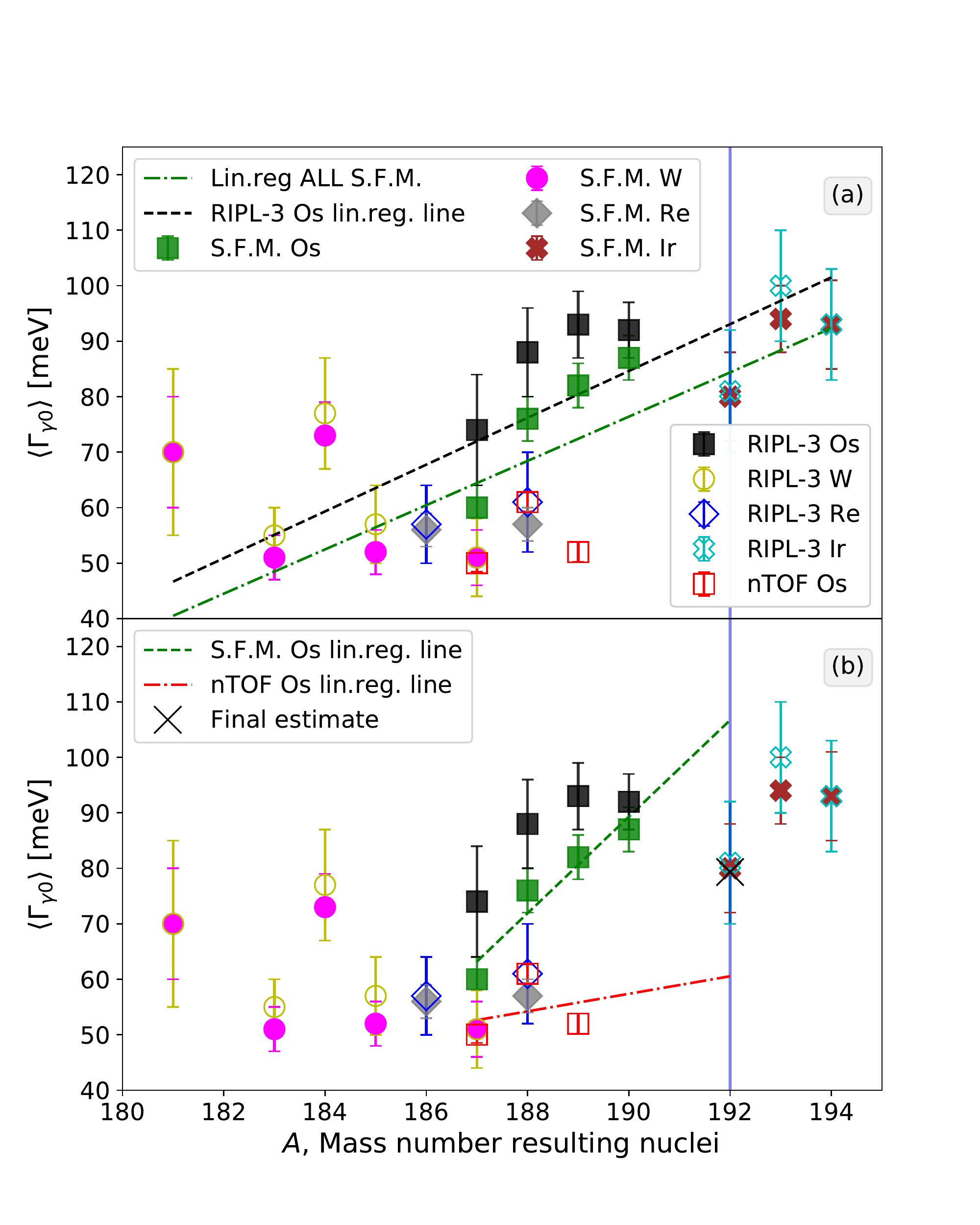}
\caption{(Color online) The average radiative widths $\langle \Gamma_{\gamma0} \rangle$ at $S_n$ plotted against mass number $A$ for osmium (squares), tungsten (circles), rhenium (diamond), and iridium (cross). The data are collected from RIPL-3 \cite{Capote2009}, Mughabghab \textit{et al.}~\cite{Mughabghab2006} (S.F.M.), and the n\_TOF Collaboration \cite{Fujii2010}. See the text for information on the linear regressions.  }
\label{fig: Gg_systematics}
\end{figure}

Experimental $\langle \Gamma_{\gamma0} \rangle$ values are not available for \Os, due to the fact that $^{191}$Os is unstable. Therefore, the $\langle \Gamma_{\gamma0} \rangle$ value at $S_n$ was estimated through multiple weighted linear regression of the values presented in \autoref{fig: Gg_systematics}, setting the weights to the inverse of the uncertainty of the points, \textit{i.e.}, a simplistic first-order estimate. The average radiative widths are plotted against mass number $A$ for available osmium isotopes in addition to the neighboring isotopes tungsten, rhenium, and iridium.  
The black dotted line is a linear regression of the RIPL-3 values, the green dash-dotted line of all the Mughabghab values (\autoref{fig: Gg_systematics}\textcolor{blue}{a}), the red dash-dotted line of the n\_TOF values, and the green dotted line of only the osmium points from Mughabghab (\autoref{fig: Gg_systematics}\textcolor{blue}{b}).

%---------------------------------------------------%
%---------------------------------------------------%
%---------------------------------------------------%

An inconsistency between the $\langle \Gamma_{\gamma0} \rangle$ values collected from RIPL-3 \cite{Capote2009}, the n\_TOF Collaboration \cite{Fujii2010}, and Mughabghab \cite{Mughabghab2006} is revealed for the osmium isotopes in \autoref{fig: Gg_systematics}. To investigate the discrepancy, a linear regression on the separate data sets was performed leading to a lower (L) and a higher (H) estimate of the $\langle \Gamma_{\gamma0} \rangle$ value for \Os\ displayed in \autoref{tab:rec_LH}. The lower estimate is calculated through an extrapolation using the n\_TOF values (red dash-dotted line, lower panel) and the higher estimate is provided by an extrapolation using the osmium values from Mughabghab \textit{et al.} (green dotted line, lower panel). 

In addition, the osmium data points display a different trend than the isotopes in the same mass region. Another estimate is therefore calculated through a linear regression of all RIPL-3 $\langle \Gamma_{\gamma0} \rangle$ values (black dotted line), and another by using all of the values provided by Mughabghab (green dash-dotted line) displayed in \autoref{fig: Gg_systematics}. The final estimate, or the ``recommended'' value (back cross) is calculated as the mean of the $\langle \Gamma_{\gamma0} \rangle$ values obtained through the four linear regressions mentioned above, \textit{i.e.}, of the points where the blue, horizontal solid line crosses the four linear regression lines. 

%---------------------------------------------------%
\begin{figure}[tb]
\centering
\includegraphics[clip,width=1.0\columnwidth]{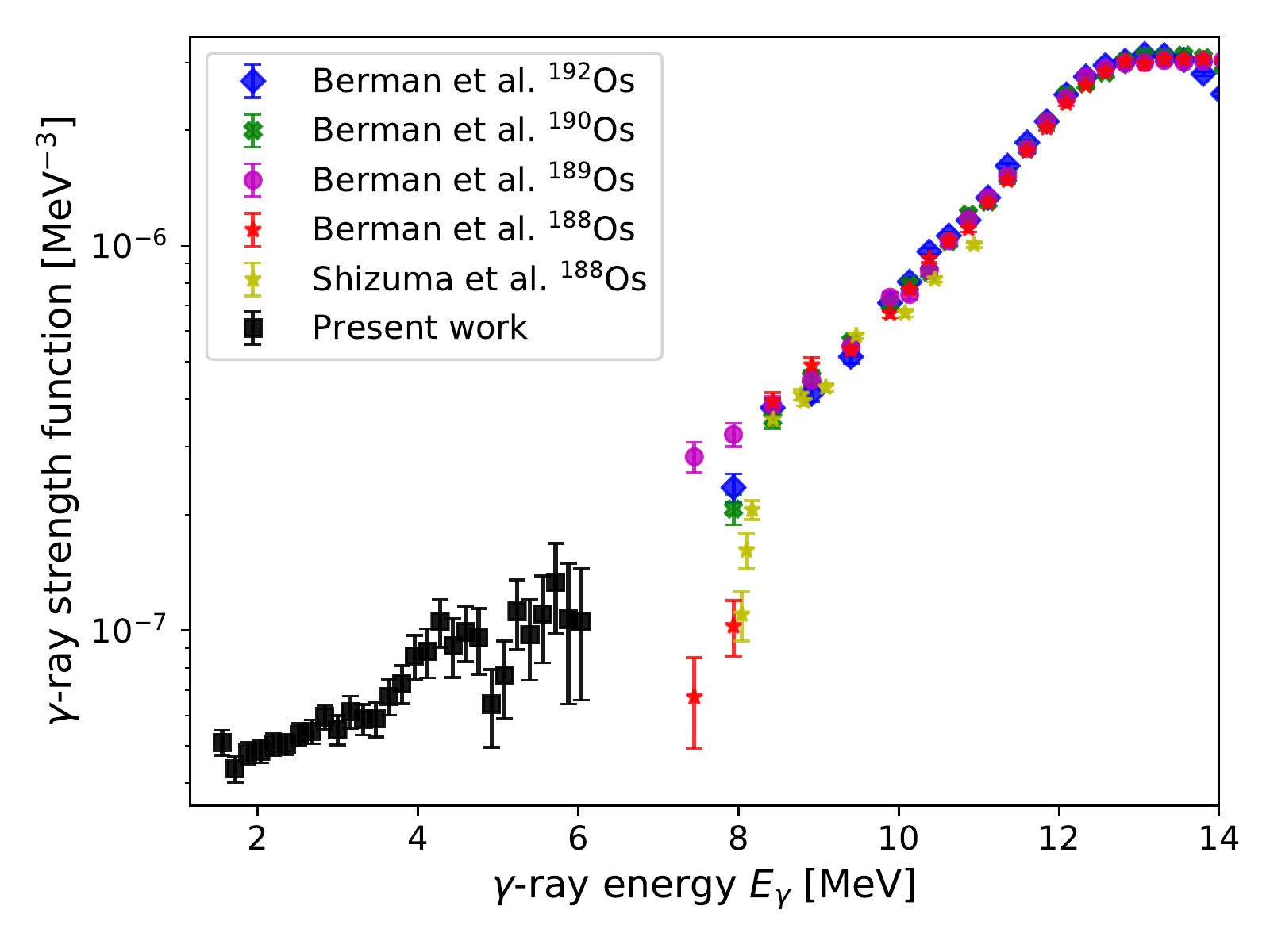}
\caption{(Color online) The $\gamma$-strength function (black) versus $\gamma$-ray energy obtained using the recommended $\langle \Gamma_{\gamma0} \rangle$ value at $S_n$. In addition, $(\gamma,n)$ data (colors) from \cite{Berman1979,Shizuma2005} are presented. Systematic errors from the normalization procedure are not included.}
\label{fig: strength_pretty}
\end{figure}
%---------------------------------------------------%
%---------------------------------------------------%

In \autoref{fig: strength_pretty}, the \ga SF obtained using the recommended $\langle \Gamma_{\gamma0} \rangle$ value at $S_n$ (see \autoref{tab:rec_LH}) is displayed together with $(\gamma,n)$ data from \cite{Berman1979}. The large error bars at high $E_\gamma$ in the  $\gamma$-strength function are directly related to the low statistics for high-energy $\gamma$ rays in \autoref{fig:coincidence}\textcolor{blue}{c}. Assuming $L=1$ and applying the principle of detailed balance~\cite{Blatt_Weisskopf}, the photonuclear cross section $\sigma_{\gamma n}$, \textit{i.e.}, the data from \cite{Berman1979,Shizuma2005}, is transformed into the \ga SF by \cite{Axel1968}
\begin{equation}
f(E_\gamma) = \frac{1}{3\pi^2 \hbar^2 c^2} \frac{\sigma_{\gamma n}(E_\gamma)}{E_\gamma}.
\end{equation}
As for the level density, the uncertainties presented in \autoref{fig: strength_pretty} are statistical errors and systematic errors of the unfolding and first-generation method \cite{Schiller2000}, \textit{i.e.}, Oslo method errors. An estimate of the systematic errors introduced by the normalization parameters will be performed in the next subsection.
  
%%%%%%%% SUBSECTION %%%%%%%%
\subsection{Systematic error estimate}
\label{sec: sys_error_est}

Often, the experimental parameters required by the normalization technique are known, and the procedure can be performed in a straightforward manner. In such cases, the uncertainty of the normalization is introduced by the model dependence of the parameter determination and the uncertainty of the experimental values applied. 
As mentioned previously, no experimental normalization parameters are known for \Os, introducing an additional uncertainty in the normalized solutions for the NLD and \ga SF. Therefore, an error estimate is of utmost importance in order to quantify the significance of the results obtained. 
 
The Oslo method software \cite{oslo-method-software} propagates the statistical and systematic errors of the analysis as described in \autoref{sec:exp_analysis}, but it does not take into account the systematic errors introduced by the choice of normalization parameters. Therefore, we estimate the total uncertainty in the following manner. The $D_0$, $\rho(S_n)$, $\sigma_I(S_n)$, and  $\langle \Gamma_{\gamma0} \rangle$ values in \autoref{tab:rec_LH} were varied to obtain a systematic uncertainty band for the NLD and \ga SF. The $D_0$ and the $\rho(S_n)$ values are correlated: using the lower $D_0$ value implies using the higher $\rho(S_n)$ value. Thus, there are only three normalization parameters to vary, \textit{i.e.}, $\sigma(S_n)$, $\langle \Gamma_{\gamma0} \rangle$, and $D_0$ or $\rho(S_n)$.

The approximated standard deviation of the level density, $\sigma_{\rho}$, due to systematic and statistical errors was split into:
\begin{align}
\sigma_{\rho,\mathrm{high}}^2 = \rho_{\mathrm{rec}}^2 \Bigg[  \bigg( \frac{ \rho_{D_0,\mathrm{low}} - \rho_{\mathrm{rec}} }{\rho_{\mathrm{rec}}} \bigg)^2 &+ \bigg( \frac{ \rho_{I,\mathrm{high}} - \rho_{\mathrm{rec}} }{\rho_{\mathrm{rec}}} \bigg)^2  \nonumber \\ 
 &+ \bigg( \frac{ \Delta \rho_{\mathrm{rec}} }{\rho_{\mathrm{rec}}} \bigg)^2 \Bigg]
\end{align}
and
\begin{align}
\sigma_{\rho,\mathrm{low}}^2 = \rho_{\mathrm{rec}}^2 \Bigg[  \bigg( \frac{ \rho_{D_0,\mathrm{high}} - \rho_{\mathrm{rec}} }{\rho_{\mathrm{rec}}} \bigg)^2 &+ \bigg( \frac{ \rho_{I,\mathrm{low}} - \rho_{\mathrm{rec}} }{\rho_{\mathrm{rec}}} \bigg)^2 \nonumber \\
&+ \bigg( \frac{ \Delta \rho_{\mathrm{rec}} }{\rho_{\mathrm{rec}}} \bigg)^2 \Bigg],
\end{align}
due to the asymmetric higher and lower estimates, compared to the recommended level density $\rho_{\mathrm{rec}}$, listed in \autoref{tab:rec_LH}. Here, $\rho_{D_0,\mathrm{low}}$ corresponds to the NLD obtained using the lower $D_0$ value (giving a high NLD) and $\rho_{D_0,\mathrm{high}}$ corresponds to the NLD obtained using the higher $D_0$ value (giving a low NLD). 
Furthermore, $\rho_{I,\mathrm{low}}$ is obtained using the low spin cutoff value $\sigma_{I,\mathrm{low}}$, and $\sigma_{I,\mathrm{high}}$ gives the $\rho_{I,\mathrm{high}}$ value (see also \autoref{tab:rec_LH}).
The Oslo method error of the recommended NLD is denoted as $\Delta \rho_{\mathrm{rec}}$.

Similarly, for the $\gamma$-strength function the standard deviation due to systematic and statistical errors $\sigma_{f}$ was estimated as
\begin{align}
\sigma_{f,\mathrm{high}}^2 = f_{\mathrm{rec}}^2 \Bigg[  \bigg( \frac{ f_{D_0,\mathrm{low}} - f_{\mathrm{rec}} }{f_{\mathrm{rec}}} \bigg)^2 + \bigg( \frac{ f_{I,\mathrm{high}} - f_{\mathrm{rec}} }{f_{\mathrm{rec}}} \bigg)^2 &\\ \nonumber
 + \bigg( \frac{ f_{ \langle \Gamma_{\gamma0} \rangle,\mathrm{high}} - f_{\mathrm{rec}} }{f_{\mathrm{rec}}} \bigg)^2 + \bigg( \frac{ \Delta f_{\mathrm{rec}} }{f_{\mathrm{rec}}} \bigg)^2 \Bigg] &
\end{align}
and
\begin{align}
\sigma_{f,\mathrm{low}}^2 = f_{\mathrm{rec}}^2 \Bigg[  \bigg( \frac{ f_{D_0,\mathrm{high}} - f_{\mathrm{rec}} }{f_{\mathrm{rec}}} \bigg)^2 + \bigg( \frac{ f_{I,\mathrm{low}} - f_{\mathrm{rec}} }{f_{\mathrm{rec}}} \bigg)^2 &\\ \nonumber
 + \bigg( \frac{ f_{ \langle \Gamma_{\gamma0} \rangle,\mathrm{low}} - f_{\mathrm{rec}} }{f_{\mathrm{rec}}} \bigg)^2 + \bigg( \frac{ \Delta f_{\mathrm{rec}} }{f_{\mathrm{rec}}} \bigg)^2 \Bigg], &
\end{align}
where $\Delta f_{\mathrm{rec}}$ is the Oslo method error of the recommended \ga SF. Hence, the higher and lower values of the level density and \ga SF  within their one-standard-deviation limits  are set to
\begin{align}
\rho_{\mathrm{high}} &= \rho_{\mathrm{rec}} + \sigma_{\rho,\mathrm{high}},\\
\rho_{\mathrm{low}} &= \rho_{\mathrm{rec}} - \sigma_{\rho,\mathrm{low}}
\end{align}
and
\begin{align}
f_{\mathrm{high}} &= f_{\mathrm{rec}} + \sigma_{f,\mathrm{high}},\\
f_{\mathrm{low}} &= f_{\mathrm{rec}} - \sigma_{f,\mathrm{low}}.
\end{align}

\begin{figure}[tb]
\centering
\includegraphics[clip,width=1.0\columnwidth]{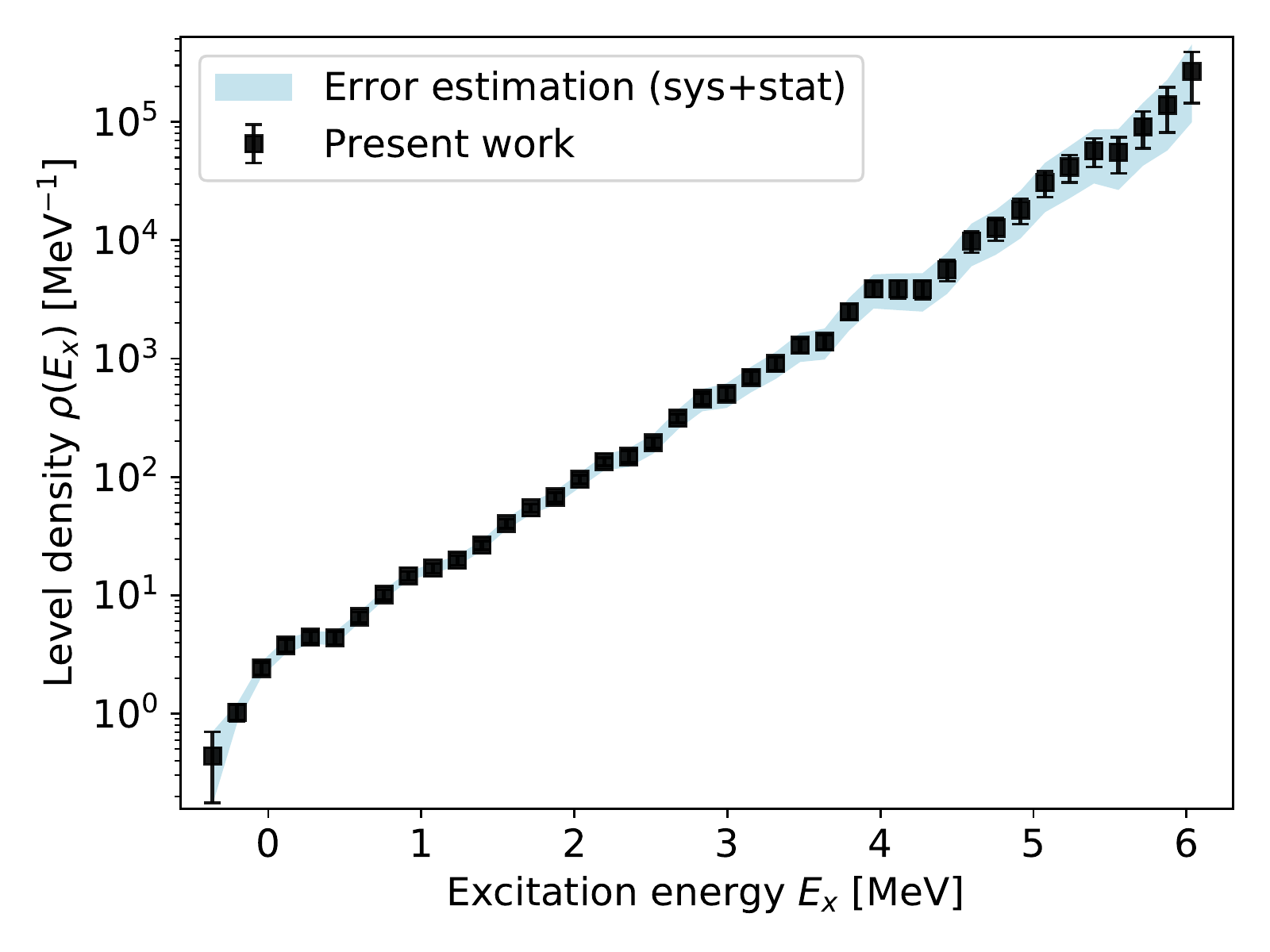}
\caption{(Color online) The level density versus excitation energy for the Oslo data (black). The Oslo method errors combined with the systematic error estimate of the normalization parameters are presented as the light-blue colored band. }
\label{fig:rho_errorband}
\end{figure}

\begin{figure}[tb]
\centering
\includegraphics[clip,width=1.0\columnwidth]{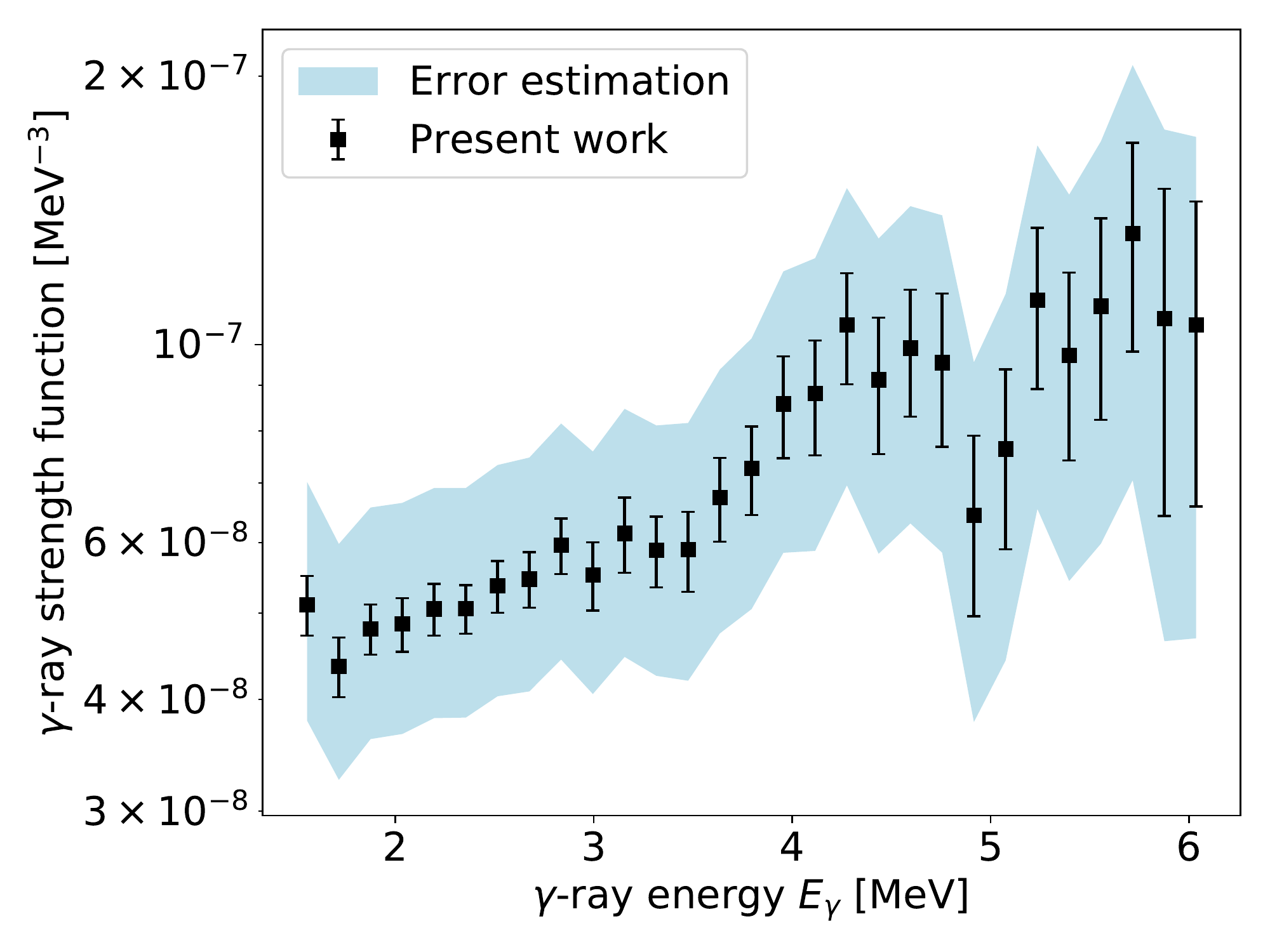}
\caption{(Color online) The $\gamma$-strength function (black) versus $\gamma$-ray energy for the Oslo data (black). The Oslo method errors combined with the systematic error estimate of the normalization parameters are presented as the light-blue colored band. }
\label{fig:strength_errorband}
\end{figure}

The resulting error band for the NLD and \ga SF of \Os\ are presented in \autoref{fig:rho_errorband} and \autoref{fig:strength_errorband}, respectively. 

%---------------------------------------------------%
%---------------------------------------------------%
%---------------------------------------------------%

%---------------------------------------------------%
%---------------------------------------------------%
%---------------------------------------------------%

%%%%%%%%%%%%%%%%%%%%%%%%%%%% SECTION %%%%%%%%%%%%%%%%%%%%%%%%%%%%%%%%%%%%%%%%%%%%%%%%

\section{Discussion of the experimental results}
\label{sec:discussion}

%%%%%%%% SUBSECTION %%%%%%%%
\subsection{The level density}
\label{subsec:nld_disc}
In \autoref{fig:rho_norm}, the nuclear level density versus excitation energy $E_x$ for \Os\ is displayed together with the CT formula (red) and the discrete known levels (blue). Up to $\approx 2$ MeV, the discrete level scheme appears to be complete, \textit{i.e.}, up to this point the discrete levels follow the same trend as the present experimental level density. An exponential growth, \textit{i.e.}, linear growth in logarithmic scale, is observed. This fits well with the CT formula, displayed in red, which has been adapted to go through the point at $\rho(S_n)$ in addition to the experimental level density. 

A quite large total uncertainty is deduced in the NLD for \Os\ in \autoref{fig:rho_errorband}; in particular the last five data points have a total uncertainty larger than 50\%. The Oslo method errors displayed in black have an uncertainty of $\approx9$\% at minimum and $\approx 45$\% at the highest-$E_x$ data point. Similarly, the average of the high and low normalization uncertainty is $\approx 12$\% at the lowest-$E_x$ data point, $\approx1$\% at minimum, and $\approx 48$\% at the highest-$E_x$ data point. The errors calculated by the Oslo method software have approximately the same magnitude and a similar evolution as the uncertainty introduced by the normalization parameters. 

%%%%%%%% SUBSECTION %%%%%%%%
\subsection{The \gsf}
\label{subsec:gsf_disc}

\begin{figure}[tb]
\centering
\includegraphics[clip,width=1.\columnwidth]{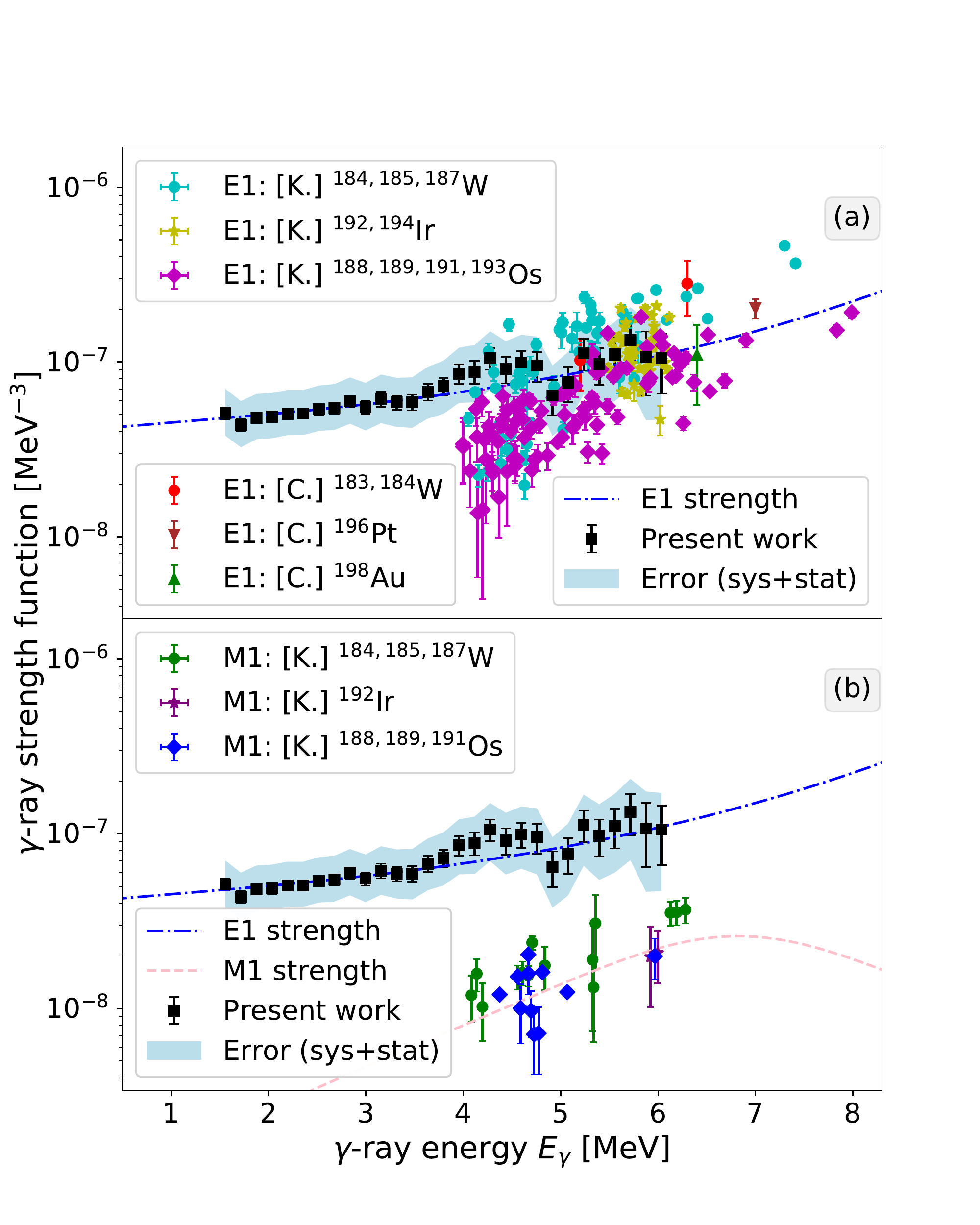}
\caption{(Color online) The $\gamma$-strength function versus $\gamma$-ray energy together with strength values of osmium (diamond), iridium (star), tungsten (circle), platinum (downward triangle), and gold (upward triangle) from Kopecky \textit{et al.}~\cite{Kopecky2017} and Capote \textit{et al.}~\cite{Capote2009}. The $E1$ (blue line) and $M1$ strengths (pink line) are estimated using a fit to the points shown in (a) and (b), respectively. The colored band represent the systematic and statistical uncertainties obtained in \autoref{sec: sys_error_est}. }
\label{fig: gsf_errorband}
\end{figure}

The \gsf\ versus \ga-ray energy for \Os\ is displayed in \autoref{fig: gsf_errorband} together with data points from \citet{Kopecky2017} (K.) and \citet{Capote2009} (C.). 

Similarly to the NLD, the \ga SF displays quite large uncertainties (blue band). The Oslo method errors, displayed in black in \autoref{fig:strength_errorband} and \ref{fig: gsf_errorband}, increase in magnitude at higher \ga-ray energy, from $\approx 8$\% at the lowest-$E_\gamma$ data point to $\approx40$\% at the highest-$E_\gamma$ data point. The average of the higher and lower normalization uncertainties has a similar magnitude over the whole energy range, varying between $\approx 31$\% at low $\gamma$-ray energy and $\approx46$\% at high $\gamma$-ray energy. Combined, the systematic and statistical uncertainty varies between $\approx 30$\% at low energy and $\approx60$\% at high energy. 

In general, the upper relative uncertainty estimate has a larger value than the lower relative estimate for the \ga SF, due to the asymmetric higher and lower estimates of the $\langle \Gamma_{\gamma0} \rangle$ value in \autoref{tab:rec_LH}. In addition, the smallest uncertainty of the \ga SF is larger than the smallest error of the NLD. This is due to the $\langle \Gamma_\gamma \rangle$ value, which only influences the uncertainty of the \ga SF. The normalization of the NLD is not dependent on the  $\langle \Gamma_\gamma \rangle$ value, while the uncertainty of the NLD normalization parameters propagates into the \ga SF error. Compared to the $(\gamma,n)$ data points by \citet{Berman1979} and \citet{Shizuma2005} in \autoref{fig: strength_pretty}, the absolute value of the \ga SF looks reasonable. 

Due to the very low statistics for  $E_\gamma > 5$ MeV and consequently large uncertainties of the present data set, it is difficult to say whether there exist any significant structures in this region of the \ga SF. 
It could be that a pygmy dipole resonance (see Ref.~\cite{Savran2013} and references therein) exists in the osmium isotopes, as observed in, \textit{e.g.}, $^{181}$Ta~\cite{Makinaga2014}. Its presence could largely affect the MACS calculations~\cite{Tsoneva2015}, but unfortunately it is not possible to conclude about the existence of such structures with the present results. 
Other experiments, possibly with a different probe than $\alpha$ particles, might shed new light on this very interesting question.

%---------------------------------------------------%
%---------------------------------------------------%
%---------------------------------------------------%

%%%%%%%% SUBSECTION %%%%%%%%
\subsection{Separating the E1 and M1 strengths}

To apply the experimental \ga SF obtained in the previous section in the cross section \textsf{TALYS} calculation, the contributions to the \ga SF from the $E1$ and $M1$ strengths have to be separated. The $E1$ and $M1$ strengths of the final \ga SF were found by a fitting $f^{\mathrm{total}}=f_{E1}^{\mathrm{GLO}}+f_{M1}^{\mathrm{SLO}}$ to the Oslo data and the $(\gamma,n)$ data points of Berman \textit{et al.}~\cite{Berman1979} for \Os\ in \autoref{fig: strength_pretty}. Here, the generalized Lorentzian (GLO) model describes the giant electric dipole resonance ($E1$) as \cite{Capote2009} 
\begin{equation}
f_{E1}^{\mathrm{GLO}} (E_\gamma) = \frac{\sigma_{E1} \Gamma_{E1} }{3\pi^2 \hbar^2c^2} \Bigg( \frac{ E_\gamma \Gamma_K}{(E_\gamma^2 - E_{E1}^2 )^2 + E_\gamma^2 \Gamma_{K}^2} + 0.7\frac{\Gamma_{K,0}}{E_{E1}^3} \Bigg),
\label{eq: GLO}
\end{equation}
where the last term is introduced to have a non zero strength in the $E_\gamma \rightarrow 0$ limit. The peak cross section, energy centroid, and width parameters are denoted as $\sigma_{E1}$, $E_{E1}$, and $\Gamma_{E1}$, respectively. Further, $\Gamma_K$ is a function of the \ga-ray energy $E_\gamma$, and the nuclear temperature parameter $T_f$ of the final levels $E_f=E_x-E_\gamma$ is given by
\begin{equation}
\Gamma_K (E_\gamma,T_f) = \frac{\Gamma_{E1}}{E_{E1}^2} (E_\gamma^2 + 4\pi^2 T_f^2),
\end{equation}
so that $\Gamma_{K,0} = \Gamma_K (0,T_f)$. The giant magnetic dipole resonance ($M1$) can be described by a standard Lorentzian (SLO) curve \cite{Capote2009}: 
\begin{equation}
f^{\mathrm{SLO}} (E_\gamma) = \frac{1}{3\pi^2 \hbar^2c^2} \frac{\sigma_{M1} E_\gamma \Gamma_{M1}^2}{(E_\gamma^2 - E_{M1}^2 )^2 + E_\gamma^2 \Gamma_{M1}^2},
\label{eq: SLO}
\end{equation}
where $\sigma_{M1}$, $E_{M1}$, and $\Gamma_{M1}$ are the peak cross section, energy centroid, and width of the SLO. A global parametrization \cite{Capote2009} provides $E_{M1} = 41\cdot A^{-1/3}$ MeV and $\Gamma_{M1} = 4$ MeV for a given mass number $A$. The peak cross section $\sigma_{M1}$ can be determined by experimental data, or by using the relation $f_{E1}/f_{M1} = 0.0588 \cdot A^{0.878}$ at $\simeq 7$ MeV \cite{Capote2009}. 

The number of free parameters for the SLO [\autoref{eq: SLO}] was reduced by using the global parametrization of \cite{Capote2009} for the SLO parameters, \textit{i.e.}, setting the width and the energy centroid constant; the only unconstrained parameter of the SLO function was the peak cross section $\sigma_{SLO}$. For the GLO parameters, on the other hand, the best fit of all parameters was found by treating all parameters as free with no constraints. The optimized parameters are listed in \autoref{tab:GLO_SLO} together with the corresponding uncertainty obtained by using scipy's curve fitting procedure \cite{SciPy}. 

%---------------------------------------------------%
\begin{table*}[tb]
\centering
\caption{The fit parameters for the GLO and SLO functions [\autoref{eq: GLO} and \ref{eq: SLO}], using the high (H), recommended (R), and lower (L) \ga SF and the $(\gamma,n)$ data points by \cite{Berman1979}.  The width $E_{M1}$ and the energy centroid $\Gamma_{M1}$ of the SLO function were held constant during the fitting procedure. The uncertainties were determined through the best fit procedure of \cite{SciPy}.}
\begin{tabular}{cccccccc} 
\hline \hline
 & $E_{E1}$ [MeV] & $\Gamma_{E1}$ [MeV] & $\sigma_{E1}$ \newline[mb] & $T_f$ [MeV] & $E_{M1}$ [MeV] & $\Gamma_{M1}$ [MeV] & $\sigma_{M1}$ \newline [mb] \\ 
\hline 
H & 13.1(1) & 2.3(2) & 713(60) & 1.5(2) & 7.1 & 4 & 4.5(4) \\ 
R & 13.2(1) & 2.8(3) & 615(43) & 1.2(2) & 7.1 & 4 & 2.1(4) \\ 
L & 13.2(1) & 3.0(3) & 572(37) & 1.0(2) & 7.1 & 4 & 0.7(4) \\ 
\hline \hline
\end{tabular}
\label{tab:GLO_SLO}
\end{table*}
%---------------------------------------------------%
%---------------------------------------------------%

The result of the fitting procedure is presented in the top and bottom panels of \autoref{fig: gsf_errorband} as the blue and pink dotted lines for the $E1$ and $M1$ strengths, respectively. As expected, the overall shape of the GLO is well adjusted to the Oslo data and the giant electric dipole resonance represented by the $(\gamma,n)$ data points (the latter are not shown in \autoref{fig: gsf_errorband}). Similarly, the absolute value of the estimated $M1$ strength agrees well with the ${}^{188,189,191}$Os, ${}^{192}$Ir, and ${}^{184,185,187}$W $M1$ values given in Ref.~\cite{Kopecky2017} [\autoref{fig: gsf_errorband}\textcolor{blue}{b}]. The listed uncertainties of the fitting procedure for the higher, recommended, and lower $\sigma_{M1}$ values are quite large compared to the uncertainties of the fitted $E1$ parameters. The $M1$ component of the strength function is orders of magnitude smaller than the $E1$ component, hence it is natural that the relative uncertainty of the fitting procedure is larger for the $M1$ component.

A large spread in strength in the $E1$ data provided by Kopecky \textit{et al.} \cite{Kopecky2017} of ${}^{192,194}$Ir, ${}^{188,189,191,193}$Os and ${}^{184,185,187}$W is observed, which can be expected to be caused by Porter-Thomas-type fluctuations~\cite{Porter1956} of  transition strengths.
Within the statistical model with a very large number of wave-function components for each level, the Porter-Thomas distribution\footnote{The Porter-Thomas distribution is a $\chi^2$ distribution with one degree of freedom.} of the partial $\gamma$-decay widths is valid, and sampling only a few transitions would lead to large fluctuations in the measured strength. 
It should be pointed out that none of the data points provided by \cite{Kopecky2017} or \cite{Capote2009} are $E1$ or $M1$ parameters of the \Os\ isotope, and they were not included in the fit. Nevertheless, these data points do provide a consistency check, \textit{i.e.}, that the experimental $E1$ and $M1$ strengths of nuclei in this mass region have approximately the same magnitude as the $E1$ and $M1$ components estimated for $^{192}$Os. 
Therefore, we deem that the  decomposition of the experimental \ga SF into the presented $E1$ and $M1$ strengths is reasonable, and it will be used to experimentally constrain the MACS value of the $^{191}$O\lowercase{s}$(n,\gamma)$ reaction in the next section.

%%%%%%%%%%%%%%%%%%%%%%%%%%%% SECTION %%%%%%%%%%%%%%%%%%%%%%%%%%%%%%%%%%%%%%%%%%%%%%%%

\section{Calculation of the $^{191}$O\lowercase{s}$(n,\gamma)$ radiative neutron capture cross section}
\label{sec:calc_ng}

We now use our results to constrain the MACS of the \Osreacto\ reaction by applying the experimental NLD and \ga SF in the \textsf{TALYS-1.9} \cite{TALYS, Koning2012} nuclear reaction code. The recommended MACS value is calculated by applying the recommended experimental NLD and \ga SF, in addition to a small set of input keywords. For instance, the keyword \texttt{Nlevels} was used to limit the number of known discrete levels, from tables such as \cite{NNDC}, used by \textsf{TALYS}. In this case the number of levels was set to 29, as this is approximately where the discrete levels saturate, \textit{i.e.}, where the experimental level density and the discrete levels show a significantly different slope at $E_x\approx2$ MeV (\autoref{fig:rho_norm}). 
In addition, a default \textsf{TALYS} calculation was done, \textit{i.e.}, a calculation where all keywords used are set to default, including the level density and \ga SF, except for the small set of keywords specified in the recommended calculation, for consistency. 

To quantify the uncertainty of the experimentally constrained MACS value, the error bands of the NLD and \ga SF are propagated into the uncertainty of the MACS values. The upper and lower limits of the NLD and \ga SF obtained in the previous section are applied to calculate the standard deviation of the MACS value in a similar manner to the standard deviation of the NLD and the \ga SF in \autoref{sec: sys_error_est}. We assumed all errors to be independent and added the combinations in quadrature. 

The OMP applied in the recommended calculation is the default \textsf{TALYS} choice: the global, phenomenological potential (\texttt{localomp n}) of Koning and Delaroche~\cite{Koning2003}. Another option for the OMP in \textsf{TALYS} is the semi microscopic, spherical Jeukenne-Lejeune-Mahaux (JLM) potential~\cite{Jeukenne1977} (\texttt{jlmomp}) as adapted by Bauge \textit{et al.}~\cite{Bauge2001}. 
To account for the uncertainty introduced by the OMP model choice, a \textsf{TALYS} calculation was done with the default potential and the JLM potential. 
The result of this calculation was then included in the uncertainty estimate. 

Another possible model choice implemented in \textsf{TALYS} is a width fluctuation correction factor (\texttt{widthmode}), where the default choice is the approximate expressions of Moldauer~\cite{Moldauer1976,Moldauer1980}. Hilaire \textit{et al.}~\cite{Hilaire2003} recommended the Moldauer expression as it better reproduced the exact expression for the width fluctuation correction factor as obtained within the Gaussian orthogonal ensemble approach.
Therefore, we have chosen to not vary the width fluctuation input keyword in the present work.

\begin{figure}[tb]
\centering
\includegraphics[clip,width=1.\columnwidth]{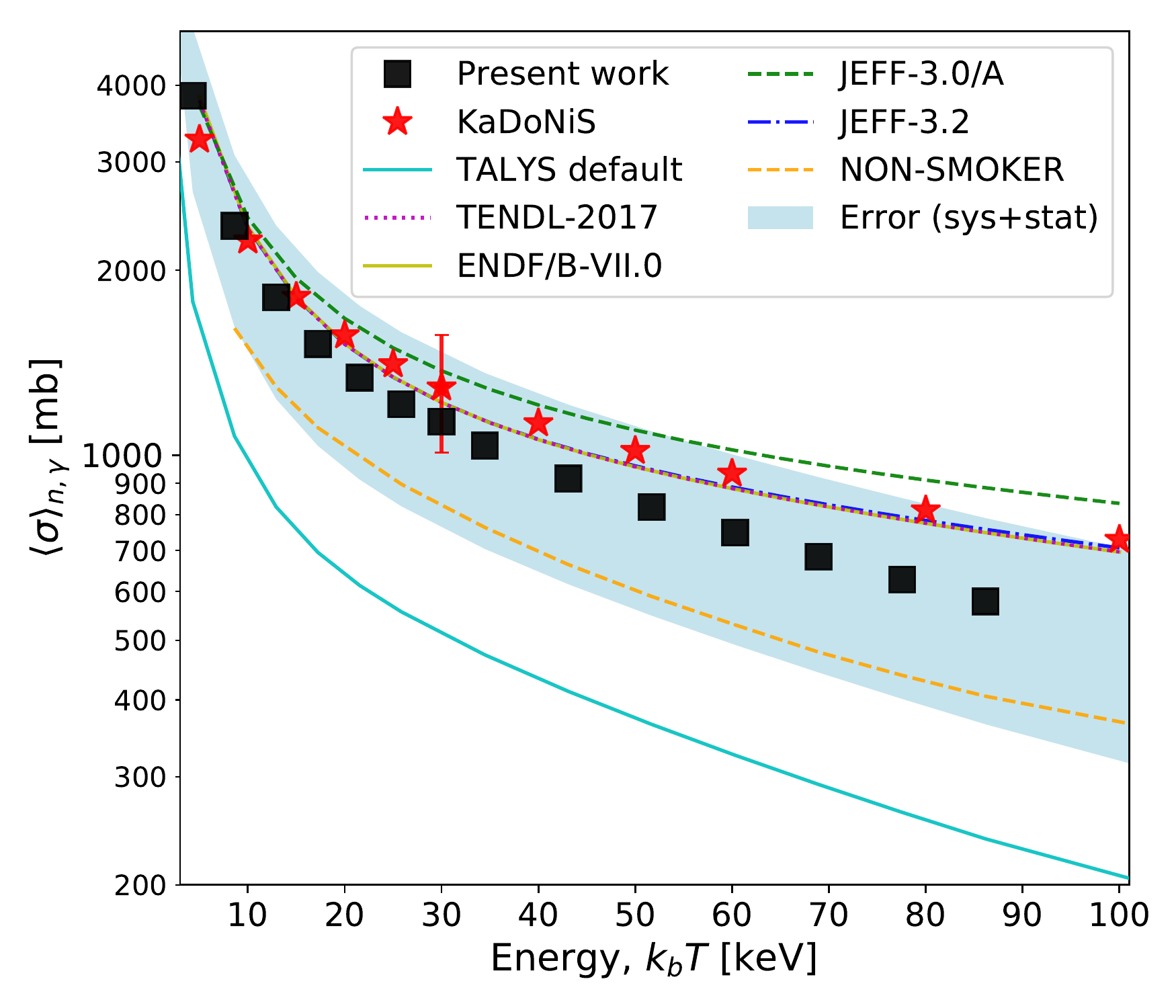}
\caption{(Color online) The Maxwellian-averaged $(n,\gamma)$ cross section versus energy $k_BT$ at the stellar temperature $T$. The systematic and statistical uncertainties (blue band) of the recommended MACS for the \Osreacto\ reaction are presented as the colored band. Values from a \textsf{TALYS} calculation done with \textit{default} input is displayed in addition to several theoretical and evaluated libraries: the KADoNiS library \cite{KadonisWeb},  TENDL-2017 \cite{Koning2012}, ENDF/B-VII.0 \cite{CHADWICK2006}, JEFF-3.0/A \cite{JEFFWeb}, JEFF-3.2 \cite{JEFFWeb}, and NON-SMOKER \cite{RaT99}. }
\label{fig:macs}
\end{figure}

The recommended Maxwellian-averaged cross section for the \Osreacto\ reaction calculated with the \textsf{TALYS} code is presented in \autoref{fig:macs} together with the default \textsf{TALYS} run, in addition to several theoretical and evaluated libraries: the KADoNiS library \cite{KadonisWeb},  TENDL-2017 \cite{Koning2012,TENDL-2017Web}, ENDF/B-VII.0 \cite{CHADWICK2006}, JEFF-3.0/A \cite{JEFFWeb}, and JEFF-3.2 \cite{JEFFWeb}.
The propagated uncertainty lies between $\approx 30$\% and $\approx 37$\% at low and high energy, respectively. Although the uncertainties estimated in the current work are asymmetrical, the upper and lower standard deviations are rather similar in magnitude, providing a mean of $\delta^{\mathrm{MACS}}_{\mathrm{mean}}=(348.8\ \mathrm{mb}+401.4\ \mathrm{mb})/2\approx375$ mb at the energy of $k_bT=30$ keV; see \autoref{tab:macs}. 
 
%---------------------------------------------------%
%---------------------------------------------------%
%---------------------------------------------------%
 
For the energy range $k_BT =5-100$ keV in \autoref{fig:macs}, the values provided by KADoNiS \cite{KadonisWeb} agree well with the MACS values obtained in this work. 
The values from the TENDL-2017, ENDF/B-VII.0, and JEFF-3.2 libraries are very similar (but not identical), and follow the same trend as the KADoNiS values. 
The theoretical NON-SMOKER values lie within the uncertainty band, as do also the JEFF-3.0/A values in the range 5-60 keV.
Interestingly, the theoretical estimate provided by the \textsf{TALYS} default input parameters gives a significantly smaller cross section than the present result and the KADoNiS values over the whole energy range. 
This is likely due to an overall lower default level density and \ga SF in the \textsf{TALYS} code compared to the present experimental data and to the models used in KADoNiS. 
The fact that using default input values gives such a low (in this case) value for the MACS demonstrates the danger of just applying default values when no experimental information is available.
We also note that although \textsf{TALYS} is applied for the TENDL-2017 library, the default inputs are not used, as stated on the TENDL-2017 web page\footnote{\url{https://tendl.web.psi.ch/tendl_2017/tendl2017.html}}: ``Not a single neutron evaluation is based on default calculations.''
Thus, although the same nuclear reaction code is the basis for our \textsf{TALYS} calculation and the TENDL-2017 value, the input NLD, $\gamma$SF, and OMP are not the same and so they are not expected to give the same output MACS.

%---------------------------------------------------%
\begin{table*}[tb]
\centering
\caption{The Maxwellian-averaged $(n,\gamma)$ cross sections with corresponding standard deviation $\delta^{\mathrm{MACS}}$ (in units of mb) at the s-process temperature $k_BT=30$ keV for the \Osreacto\ reaction. Values from KADoNiS \cite{KadonisWeb}, the \textsf{TALYS} \cite{Goriely2008} \textit{default} run, NON-SMOKER \cite{RaT99}, TENDL-2017 \cite{Koning2012}, ENDF/B-VII.0 \cite{CHADWICK2006}, JEFF-3.0/A and JEFF-3.2 \cite{JEFFWeb} are displayed.}
\begin{tabular}{ccccccccc} 
\hline \hline
   & Present work & KADoNiS & TALYS \textit{default} & NON-SMOKER & TENDL-2017 & ENDF/B-VII.0 & JEFF-3.0/A & JEFF-3.2 \\ 
\hline 
$\langle \sigma \rangle_{n,\gamma}$ & 1134 & 1290 & 523 & 802 & 1218 & 1219 & 1372 & 1219 \\ 
$\delta_{\mathrm{MACS}}$ & 375 & 280 & - & - & - & - & - & - \\
\hline \hline
\end{tabular} 
\label{tab:macs}
\end{table*}
%---------------------------------------------------%
%---------------------------------------------------%

A comparison of the MACS values at the $s$-process temperature of $k_BT=30$ keV is presented in \autoref{tab:macs}. 
As noticed before, the \textsf{TALYS} default value is significantly lower than the present result and the KADoNiS MACS, and lies outside of the error band. 
In addition, five MACS values from NON-SMOKER \cite{RaT99}, TENDL-2017 \cite{Koning2012}, ENDF/B-VII.0 \cite{CHADWICK2006}, JEFF-3.0/A, and JEFF-3.2 \cite{JEFFWeb} are presented.
The NON-SMOKER value deviates significantly from the other four libraries, but lies within the uncertainty of the present result. 

The magnitude of the systematic and statistical uncertainties leads to a rather large relative uncertainty in the present result of $\approx 33$\%, in contrast to $\approx 22$\% for the theoretical KADoNiS cross sections. 
Nevertheless, it is rewarding to observe that the theoretical value provided by KADoNiS is consistent and well within the uncertainty of the experimentally constrained MACS obtained in this work. 

%%%%%%%%%%%%%%%%%%%%%%%%%%%% SECTION %%%%%%%%%%%%%%%%%%%%%%%%%%%%%%%%%%%%%%%%%%%%%%%%

\section{Summary and outlook}
\label{sec:sum}

In this work, we have obtained the first experimentally constrained Maxwellian-averaged cross section for the \Osreacto\ reaction relevant to the s-process nucleosynthesis. The nuclear level density and the \gsf\ of \Os\ have been extracted from $\alpha-\gamma$ coincidence data using the Oslo method and normalized by means of parameters estimated from neutron-resonance data of isotopes in the vicinity of \Os. An estimation of the uncertainties introduced by the normalization parameters was done, and the resulting error band was propagated into the calculated MACS values. 

No significant structures are revealed in the nuclear level density of \Os\, and the exponential growth of the constant temperature model describes the level density well. 
The uncertainties of the \ga SF are slightly larger than the errors of the level density ($\approx 30-60$\%), but the overall absolute value agrees well with $(\gamma,n)$ data and $E1$ and $M1$ strengths provided by external data. 
The poor statistics at $E_\gamma > 5$ MeV and the correspondingly large uncertainties in the \ga SF in this region make it difficult to conclude whether a pygmy dipole resonance is present or not.
More information from experiments with different probes would be highly desirable to address this issue.

The experimentally constrained MACS, $\langle \sigma \rangle_{n,\gamma} = 1134 \pm 375$ mb at $k_BT=30$ keV, is  found to be fully consistent with the predictions listed in the KADoNiS library. Good agreement is also found with MACS values obtained from the TENDL-2017, ENDF/B-VII.0, and JEFF libraries. 
The \textsf{TALYS} default calculation, on the other hand, yields a $\approx 50$\% lower MACS value compared to the present result, lying outside of the experimental error bar.

As the present MACS presented in \autoref{fig:macs} is very similar to the KADoNiS values, using our result in an $s$-process nucleosynthesis network calculation would not change the resulting abundances significantly.
Nevertheless, we intend to perform a full $s$-process network calculation when the data analysis of the remaining branch point isotopes $^{185}$W, $^{186}$Re, and $^{192}$Ir is completed.

The extracted NLD and \ga SF data, as well as  the calculated cross section are available in the Supplemental Material~\cite{SuplementalMaterial}.

\acknowledgments

This work was financed through ERC-STG-2014 under Grant Agreement No. 637686. A.C.L. acknowledges support from the ChETEC COST Action (CA16117), supported by COST (European Cooperation in Science and Technology) and A.G. acknowledges support from the Norwegian Research Council Grant No. 263030. This work was supported in part by the National Science Foundation under Grant No. PHY-1430152 (JINA Center for the Evolution of the Elements). The authors would like to thank J. C. M\"{uller}, P. A. Sobas, and J. C. Wikne at the Oslo Cyclotron Laboratory for operating the cyclotron and providing excellent experimental conditions.

\bibliography{Oslo_Method.bib,web_references.bib,other_ref.bib,GeneralNuclearPhysics.bib,NuclearAstro.bib}

\end{document}